\documentclass[%
 reprint,
 amsmath,amssymb,
 aps,
 prd,
floatfix, 
]{revtex4-2}

\usepackage{comment}

\usepackage{graphicx}
\usepackage{dcolumn}
\usepackage{bm}
\usepackage{xcolor}
\usepackage{pifont}
\usepackage{booktabs}
\usepackage{multirow}
\usepackage{relsize}
\usepackage{algorithm}
\usepackage{algpseudocode}
\usepackage{amsmath}
\usepackage[mathlines]{lineno}

\newcommand{\GWtwothreezerofivetwonineprimarybetalowNSEOSminPNS}{\ensuremath{0}}
\newcommand{\GWtwothreezerofivetwonineprimarybetalowNSEOSmaxPNS}{\ensuremath{18}}

\newcommand{\GWoneninezeroeightonefoursecondarybetalowNSEOSminPNS}{\ensuremath{7}}
\newcommand{\GWoneninezeroeightonefoursecondarybetalowNSEOSmaxPNS}{\ensuremath{10}}

\newcommand{\GWoneninezerofourtwofiveprimarybetalowNSEOSdeltapercent}{\ensuremath{37}}
\newcommand{\GWoneninezerofourtwofiveprimarybetalowNSEOSminPNS}{\ensuremath{60}}
\newcommand{\GWoneninezerofourtwofiveprimarybetalowNSEOSmaxPNS}{\ensuremath{98}}

\newcommand{\GWtwothreezerofivetwonineprimarysigpeakNSEOSminPNS}{\ensuremath{2}}
\newcommand{\GWtwothreezerofivetwonineprimarysigpeakNSEOSmaxPNS}{\ensuremath{6}}

\newcommand{\GWoneninezeroeightonefoursecondarysigpeakNSEOSminPNS}{\ensuremath{7}}
\newcommand{\GWoneninezeroeightonefoursecondarysigpeakNSEOSmaxPNS}{\ensuremath{9}}

\newcommand{\GWoneninezerofourtwofiveprimarysigpeakNSEOSminPNS}{\ensuremath{67}}
\newcommand{\GWoneninezerofourtwofiveprimarysigpeakNSEOSmaxPNS}{\ensuremath{96}}

\newcommand{\GWtwothreezerofivetwonineprimarygammalowEOSminPNS}{\ensuremath{4}}
\newcommand{\GWtwothreezerofivetwonineprimarygammalowEOSmaxPNS}{\ensuremath{7}}

\newcommand{\GWoneninezeroeightonefoursecondarygammalowEOSminPNS}{\ensuremath{6}}
\newcommand{\GWoneninezeroeightonefoursecondarygammalowEOSmaxPNS}{\ensuremath{10}}

\newcommand{\GWoneninezerofourtwofiveprimarygammalowEOSminPNS}{\ensuremath{87}}
\newcommand{\GWoneninezerofourtwofiveprimarygammalowEOSmaxPNS}{\ensuremath{91}}

\newcommand{\GWtwothreezerofivetwonineprimarymucostiltdeltapercent}{\ensuremath{43}}
\newcommand{\GWtwothreezerofivetwonineprimarymucostiltminPNS}{\ensuremath{5}}
\newcommand{\GWtwothreezerofivetwonineprimarymucostiltmaxPNS}{\ensuremath{48}}

\newcommand{\GWoneninezeroeightonefoursecondarymucostiltminPNS}{\ensuremath{68}}
\newcommand{\GWoneninezeroeightonefoursecondarymucostiltmaxPNS}{\ensuremath{75}}

\newcommand{\GWoneninezerofourtwofiveprimarymucostiltminPNS}{\ensuremath{95}}
\newcommand{\GWoneninezerofourtwofiveprimarymucostiltmaxPNS}{\ensuremath{99}}

\newcommand{\GWtwothreezerofivetwonineprimarygammahighEOSminPNS}{\ensuremath{4}}
\newcommand{\GWtwothreezerofivetwonineprimarygammahighEOSmaxPNS}{\ensuremath{5}}

\newcommand{\GWoneninezeroeightonefoursecondarygammahighEOSminPNS}{\ensuremath{7}}
\newcommand{\GWoneninezeroeightonefoursecondarygammahighEOSmaxPNS}{\ensuremath{9}}

\newcommand{\GWoneninezerofourtwofiveprimarygammahighEOSminPNS}{\ensuremath{87}}
\newcommand{\GWoneninezerofourtwofiveprimarygammahighEOSmaxPNS}{\ensuremath{90}}

\newcommand{\GWtwothreezerofivetwonineprimarybetalowNSdeltapercent}{\ensuremath{46}}
\newcommand{\GWtwothreezerofivetwonineprimarybetalowNSminPNS}{\ensuremath{6}}
\newcommand{\GWtwothreezerofivetwonineprimarybetalowNSmaxPNS}{\ensuremath{53}}

\newcommand{\GWoneninezeroeightonefoursecondarybetalowNSminPNS}{\ensuremath{73}}
\newcommand{\GWoneninezeroeightonefoursecondarybetalowNSmaxPNS}{\ensuremath{73}}

\newcommand{\GWoneninezerofourtwofiveprimarybetalowNSminPNS}{\ensuremath{90}}
\newcommand{\GWoneninezerofourtwofiveprimarybetalowNSmaxPNS}{\ensuremath{99}}

\newcommand{\GWtwothreezerofivetwonineprimarymuchioneEOSminPNS}{\ensuremath{0}}
\newcommand{\GWtwothreezerofivetwonineprimarymuchioneEOSmaxPNS}{\ensuremath{20}}

\newcommand{\GWoneninezeroeightonefoursecondarymuchioneEOSminPNS}{\ensuremath{6}}
\newcommand{\GWoneninezeroeightonefoursecondarymuchioneEOSmaxPNS}{\ensuremath{12}}

\newcommand{\GWoneninezerofourtwofiveprimarymuchioneEOSminPNS}{\ensuremath{87}}
\newcommand{\GWoneninezerofourtwofiveprimarymuchioneEOSmaxPNS}{\ensuremath{97}}

\newcommand{\GWtwothreezerofivetwonineprimarymucostiltEOSminPNS}{\ensuremath{0}}
\newcommand{\GWtwothreezerofivetwonineprimarymucostiltEOSmaxPNS}{\ensuremath{9}}

\newcommand{\GWoneninezeroeightonefoursecondarymucostiltEOSminPNS}{\ensuremath{3}}
\newcommand{\GWoneninezeroeightonefoursecondarymucostiltEOSmaxPNS}{\ensuremath{12}}

\newcommand{\GWoneninezerofourtwofiveprimarymucostiltEOSminPNS}{\ensuremath{80}}
\newcommand{\GWoneninezerofourtwofiveprimarymucostiltEOSmaxPNS}{\ensuremath{97}}

\newcommand{\GWtwothreezerofivetwonineprimarymupeakNSEOSminPNS}{\ensuremath{3}}
\newcommand{\GWtwothreezerofivetwonineprimarymupeakNSEOSmaxPNS}{\ensuremath{9}}

\newcommand{\GWoneninezeroeightonefoursecondarymupeakNSEOSminPNS}{\ensuremath{7}}
\newcommand{\GWoneninezeroeightonefoursecondarymupeakNSEOSmaxPNS}{\ensuremath{9}}

\newcommand{\GWoneninezerofourtwofiveprimarymupeakNSEOSminPNS}{\ensuremath{84}}
\newcommand{\GWoneninezerofourtwofiveprimarymupeakNSEOSmaxPNS}{\ensuremath{90}}

\newcommand{\GWtwothreezerofivetwonineprimarysigpeakNSminPNS}{\ensuremath{18}}
\newcommand{\GWtwothreezerofivetwonineprimarysigpeakNSmaxPNS}{\ensuremath{37}}

\newcommand{\GWoneninezeroeightonefoursecondarysigpeakNSminPNS}{\ensuremath{72}}
\newcommand{\GWoneninezeroeightonefoursecondarysigpeakNSmaxPNS}{\ensuremath{73}}

\newcommand{\GWoneninezerofourtwofiveprimarysigpeakNSminPNS}{\ensuremath{93}}
\newcommand{\GWoneninezerofourtwofiveprimarysigpeakNSmaxPNS}{\ensuremath{99}}

\newcommand{\GWtwothreezerofivetwonineprimarygammahighminPNS}{\ensuremath{31}}
\newcommand{\GWtwothreezerofivetwonineprimarygammahighmaxPNS}{\ensuremath{32}}

\newcommand{\GWoneninezeroeightonefoursecondarygammahighminPNS}{\ensuremath{73}}
\newcommand{\GWoneninezeroeightonefoursecondarygammahighmaxPNS}{\ensuremath{73}}

\newcommand{\GWoneninezerofourtwofiveprimarygammahighminPNS}{\ensuremath{97}}
\newcommand{\GWoneninezerofourtwofiveprimarygammahighmaxPNS}{\ensuremath{97}}

\newcommand{\GWtwothreezerofivetwonineprimarysigmachiminPNS}{\ensuremath{18}}
\newcommand{\GWtwothreezerofivetwonineprimarysigmachimaxPNS}{\ensuremath{35}}

\newcommand{\GWoneninezeroeightonefoursecondarysigmachiminPNS}{\ensuremath{74}}
\newcommand{\GWoneninezeroeightonefoursecondarysigmachimaxPNS}{\ensuremath{74}}

\newcommand{\GWoneninezerofourtwofiveprimarysigmachiminPNS}{\ensuremath{97}}
\newcommand{\GWoneninezerofourtwofiveprimarysigmachimaxPNS}{\ensuremath{100}}

\newcommand{\GWtwothreezerofivetwonineprimarymupeakNSminPNS}{\ensuremath{26}}
\newcommand{\GWtwothreezerofivetwonineprimarymupeakNSmaxPNS}{\ensuremath{46}}

\newcommand{\GWoneninezeroeightonefoursecondarymupeakNSminPNS}{\ensuremath{72}}
\newcommand{\GWoneninezeroeightonefoursecondarymupeakNSmaxPNS}{\ensuremath{73}}

\newcommand{\GWoneninezerofourtwofiveprimarymupeakNSminPNS}{\ensuremath{96}}
\newcommand{\GWoneninezerofourtwofiveprimarymupeakNSmaxPNS}{\ensuremath{97}}

\newcommand{\GWtwothreezerofivetwonineprimarysigmachiEOSminPNS}{\ensuremath{0}}
\newcommand{\GWtwothreezerofivetwonineprimarysigmachiEOSmaxPNS}{\ensuremath{7}}

\newcommand{\GWoneninezeroeightonefoursecondarysigmachiEOSminPNS}{\ensuremath{6}}
\newcommand{\GWoneninezeroeightonefoursecondarysigmachiEOSmaxPNS}{\ensuremath{10}}

\newcommand{\GWoneninezerofourtwofiveprimarysigmachiEOSminPNS}{\ensuremath{87}}
\newcommand{\GWoneninezerofourtwofiveprimarysigmachiEOSmaxPNS}{\ensuremath{100}}

\newcommand{\GWtwothreezerofivetwonineprimarymuchioneminPNS}{\ensuremath{18}}
\newcommand{\GWtwothreezerofivetwonineprimarymuchionemaxPNS}{\ensuremath{57}}

\newcommand{\GWoneninezeroeightonefoursecondarymuchioneminPNS}{\ensuremath{72}}
\newcommand{\GWoneninezeroeightonefoursecondarymuchionemaxPNS}{\ensuremath{73}}

\newcommand{\GWoneninezerofourtwofiveprimarymuchioneminPNS}{\ensuremath{97}}
\newcommand{\GWoneninezerofourtwofiveprimarymuchionemaxPNS}{\ensuremath{99}}
\newcommand{\GWoneninezerofourtwofiveonebetamutiltcombinedmin}{\ensuremath{88}}
\newcommand{\GWoneninezerofourtwofiveonebetamutiltcombinedmax}{\ensuremath{100}}
\newcommand{\GWoneninezeroeightonefourtwobetamutiltcombinedmin}{\ensuremath{68}}
\newcommand{\GWoneninezeroeightonefourtwobetamutiltcombinedmax}{\ensuremath{75}}
\newcommand{\GWtwothreezerofivetwonineonebetamutiltcombinedmin}{\ensuremath{1}}
\newcommand{\GWtwothreezerofivetwonineonebetamutiltcombinedmax}{\ensuremath{67}}
\newcommand{\GWoneninezerofourtwofiveonebetamutiltcombinedEOSmin}{\ensuremath{51}}
\newcommand{\GWoneninezerofourtwofiveonebetamutiltcombinedEOSmax}{\ensuremath{100}}
\newcommand{\GWoneninezeroeightonefourtwobetamutiltcombinedEOSmin}{\ensuremath{4}}

\newcommand{\GWtwothreezerofivetwonineonebetamutiltcombinedEOSmin}{\ensuremath{0}}
\newcommand{\GWtwothreezerofivetwonineonebetamutiltcombinedEOSmax}{\ensuremath{21}}
\newcommand{\betaonemin}{\ensuremath{-2.91}}
 \newcommand{\betaonemax}{\ensuremath{4.60}}
 \newcommand{\betaonePNSmin}{\ensuremath{-5.00}}
 \newcommand{\betaonePNSmax}{\ensuremath{5.00}}
 \newcommand{\mupeakNSmin}{\ensuremath{1.04}}
 \newcommand{\mupeakNSmax}{\ensuremath{2.02}}
 \newcommand{\mupeakNSPNSmin}{\ensuremath{1.00}}
 \newcommand{\mupeakNSPNSmax}{\ensuremath{2.40}}
 \newcommand{\sigpeakNSmin}{\ensuremath{0.30}}
 \newcommand{\sigpeakNSmax}{\ensuremath{0.95}}
 \newcommand{\sigpeakNSPNSmin}{\ensuremath{0.00}}
 \newcommand{\sigpeakNSPNSmax}{\ensuremath{1.00}}
 \newcommand{\mucostiltmin}{\ensuremath{1.00}}
 \newcommand{\mucostiltmax}{\ensuremath{1.00}}
 \newcommand{\mucostiltPNSmin}{\ensuremath{-1.00}}
 \newcommand{\mucostiltPNSmax}{\ensuremath{1.00}}
 \newcommand{\sigtiltonemin}{\ensuremath{0.24}}
 \newcommand{\sigtiltonemax}{\ensuremath{3.76}}
 \newcommand{\sigtiltonePNSmin}{\ensuremath{0.00}}
 \newcommand{\sigtiltonePNSmax}{\ensuremath{3.00}}
 \newcommand{\muchionemin}{\ensuremath{0.01}}
 \newcommand{\muchionemax}{\ensuremath{0.38}}
 \newcommand{\muchionePNSmin}{\ensuremath{0.00}}
 \newcommand{\muchionePNSmax}{\ensuremath{1.00}}
 \newcommand{\sigchionemin}{\ensuremath{0.09}}
 \newcommand{\sigchionemax}{\ensuremath{1.90}}
 \newcommand{\sigchionePNSmin}{\ensuremath{0.00}}
 \newcommand{\sigchionePNSmax}{\ensuremath{2.00}}
 \newcommand{\gammahighmin}{\ensuremath{4.20}}
 \newcommand{\gammahighmax}{\ensuremath{7.80}}
 \newcommand{\gammahighPNSmin}{\ensuremath{4.00}}
 \newcommand{\gammahighPNSmax}{\ensuremath{8.00}}

\newcommand{\MMMSDefaultPsrGwHighSpinPrimaryIsNS}{\ensuremath{2.9 \pm 0.4\%}}
\newcommand{\MMMSDefaultPsrGwHighSpinSecondaryIsNS}{\ensuremath{96.1 \pm 0.4\%}}
\newcommand{\MMMSPDBGWTCPsrGWPDBSpinPrimaryIsNS}{\ensuremath{8.8 \pm 2.8\%}}
\newcommand{\MMMSPDBGWTCPsrGWPDBSpinSecondaryIsNS}{\ensuremath{98.4 \pm 1.3\%}}

\newcommand{\updatedMMMSDefaultPsrGwHighSpinPrimaryIsNS}{\ensuremath{3.0 \pm 0.2\%}}
\newcommand{\updatedMMMSDefaultPsrGwHighSpinSecondaryIsNS}{\ensuremath{96.2 \pm 0.2\%}}

\newcommand{\PDBsc}{\textsc{Power law + dip + break}}
\newcommand{\q}{\ensuremath{q}}
\newcommand{\gridscale}{0.99}

\newcommand{\shortpara}{\textbf}
\newcommand{\shededregionmessage}{The shaded region represents 90\% credible interval around the mean. Primaries are solid lines and secondaries are dotted lines.}
\newcommand{\topbottompopeosmessage}{(\textbf{Top}) Population-only, $P(m<\mmaxpop)$. (\textbf{Bottom}) EOS-informed, $P(m<\mmaxeos)$.}
\newcommand{\Xeff}{\ensuremath{\chi_{\text{eff}}}}
\newcommand{\mone}{\ensuremath{m_1}}
\newcommand{\mtwo}{\ensuremath{m_2}}
\newcommand{\toinclude}[1]{\textbf{\textcolor{red}{#1}}} 
\newcommand{\Msun}{\ensuremath{M_\odot}}
\newcommand{\utty}[1]{{\textcolor{black}{#1}}}  
\newcommand{\EOS}{\mathrm{EOS}}
\newcommand{\PNS}{$P(\text{NS})$}
\newcommand{\mbrk}{\ensuremath{m_{\text{break}}}}
\newcommand{\gammalow}{\ensuremath{\gamma_{\text{low}}}}
\newcommand{\gammahigh}{\ensuremath{\gamma_{\text{high}}}}
\newcommand{\etalow}{\ensuremath{\eta_{\text{low}}}}
\newcommand{\etahigh}{\ensuremath{\eta_{\text{high}}}}

\newcommand{\mtovdet}{\ensuremath{m^{\mathrm{ det}}_{\text{max}}}}
\newcommand{\mupeakNS}{\ensuremath{\mu_{\rm peak}^{\rm NS}}}
\newcommand{\sigpeakNS}{\ensuremath{\sigma_{\rm peak}^{\rm NS}}}
\newcommand{\pairingBNS}
{\ensuremath{\beta_{\rm LL}}}
\newcommand{\pairingBBH}{\ensuremath{\beta_{\rm HH}}}
\newcommand{\pairingNSBH}{\ensuremath{\beta_{\rm LH}}}

\newcommand{\mutilt}{\ensuremath{\mu^{\mathrm{low}}_{\mathrm{cos}\theta}}}
\newcommand{\sigtilt}{\ensuremath{\sigma^{\mathrm{low}}_{\mathrm{cos}\theta}}}
\newcommand{\muchi}{\ensuremath{\mu^{\mathrm{low}}_{\chi}}}
\newcommand{\sigchi}{\ensuremath{\sigma^{\mathrm{low}}_{\chi}}}
\newcommand{\mmaxpop}{\ensuremath{\gamma_{\text{low}}}}
\newcommand{\mmaxeos}{\ensuremath{m^{\text{EOS}}_{\text{max}}}}
\newcommand{\mmaxnonparam}{\ensuremath{m_{\max}^{\rm det}}}
\newcommand{\popmodel}{\textsc{FullPop-4.0}}
\newcommand{\TableSize}[1]{\normalsize{#1}}
\newcommand{\tickone}{\textcolor[rgb]{0.9216,0.8235,0.2039}{\LARGE{\ding{52}}}}   
\newcommand{\ticktwo}{\textcolor[rgb]{0.5412,0.7608,0.5020}{\LARGE{\ding{52}\ding{52}}}}   
\newcommand{\tickthree}{\textcolor[rgb]{0.0,0.6,0.0}{\LARGE{\ding{52}\ding{52}\ding{52}}}} 
\newcommand{\cross}{\textcolor[rgb]{0.8,0.0,0.0}{\huge{\ding{55}}}}     
\newcommand{\tickonetxt}{\textcolor[rgb]{0.9216,0.8235,0.2039}{\normalsize{\ding{52}}}}

\newcommand{\tickthreetxt}{\textcolor[rgb]{0.0,0.6,0.0}{\normalsize{\ding{52}\ding{52}\ding{52}}}}
\newcommand{\crosstxt}{\textcolor[rgb]{0.8,0.0,0.0}{\large{\ding{55}}}}
\newcommand{\poplabel}{Population}
\newcommand{\eoslabel}{EOS}
\newcommand{\qXeff}{\ensuremath{\q-\Xeff}}
\newcommand{\qmone}{\ensuremath{\q-\mone}}
\newcommand{\qmtwo}{\ensuremath{\q-\mtwo}}
\newcommand{\msb}{\ensuremath{m_{\mathrm{break,spin}}}}                 
\newcommand{\amin}{\ensuremath{\chi_{\min}}}                        
\newcommand{\amax}{\ensuremath{\chi_{\max}^{\mathrm{high}}}}                         
\newcommand{\amaxNS}{\ensuremath{\chi_{\max}^{\mathrm{low}}}}        
\newcommand{\ctmin}{\ensuremath{c_{\min}}}                       
\newcommand{\ctmax}{\ensuremath{c_{\max}}}                       
\newcommand{\aone}{\ensuremath{\chi_1}}
\newcommand{\atwo}{\ensuremath{\chi_2}}
\newcommand{\costiltone}{\ensuremath{\cos\theta_1}}
\newcommand{\costilttwo}{\ensuremath{\cos\theta_2}}
\newcommand{\muchiOne}{\ensuremath{\mu_{\chi}^{\rm low}}}
\newcommand{\sigchiOne}{\ensuremath{\sigma_{\chi}^{\rm low}}}
\newcommand{\muchiTwo}{\ensuremath{\mu_{\chi}^{\rm high}}}
\newcommand{\sigchiTwo}{\ensuremath{\sigma_{\chi}^{\rm high}}}
\newcommand{\mixtiltOne}{\ensuremath{\lambda^{\rm low}_{\mathrm{cos}\theta}}}
\newcommand{\sigtiltOne}{\ensuremath{\sigma^{\rm low}_{\mathrm{cos}\theta}}}
\newcommand{\mixtiltTwo}{\ensuremath{\lambda^{\rm high}_{\mathrm{cos}\theta}}}
\newcommand{\sigtiltTwo}{\ensuremath{\sigma^{\rm high}_{\mathrm{cos}\theta}}}
\newcommand{\TN}[1]{\mathcal{N}_{#1}}                             

\newcommand{\z}{\ensuremath{z}}                
\newcommand{\kappaZ}{\ensuremath{\kappa}}      
\newcommand{\resultpopgwtwothreezerofiveonine}{$\GWtwothreezerofivetwonineonebetamutiltcombinedmin$\% -- $\GWtwothreezerofivetwonineonebetamutiltcombinedmax$\%} 
\newcommand{\resulteosgwonenineofourtwofive}{$\GWoneninezerofourtwofiveonebetamutiltcombinedEOSmin$\% -- $\GWoneninezerofourtwofiveonebetamutiltcombinedEOSmax$\%} 
\newcommand{\numevents}{66}

\begin{document}
\AtBeginEnvironment{algorithm}{\nolinenumbers}
\preprint{APS/123-QED}

\title{
Guesswork in the gap: 
the impact of uncertainty in the compact binary population on source classification
}

\author{
Utkarsh Mali\(^{1}\) and Reed Essick\(^{1,2}\)
}

\affiliation{\(^{1}\)Canadian Institute for Theoretical Astrophysics and Department of Physics, University of Toronto, 60 St. George St, Toronto, ON M5S 3H8, Canada}
\affiliation{\(^{2}\)David A. Dunlap Department of Astronomy, University of Toronto, 60 St. George St, Toronto, ON M5S 3H8, Canada}

\date{\today}

\begin{abstract}
The nature of the compact objects within the supposed ``lower mass gap'' remains uncertain. 
Observations of GW190814 and GW230529 highlight the challenges gravitational waves face in distinguishing neutron stars from black holes. 
Interpreting these systems is especially difficult because classifications depend simultaneously on measurement noise, compact binary population models, and equation of state (EOS) constraints on the maximum neutron star mass. 
We analyze~\numevents\ confident events from GWTC-3 to quantify how the probability of a component being a neutron star,~\PNS, varies across the population. 
The effects are substantial, the dominant drivers of classification are the pairing preferences of neutron stars with other compact objects, and the neutron star spin distributions.
The data reveals that~\PNS\ varies between~\resultpopgwtwothreezerofiveonine\ for GW230529's primary and between \resulteosgwonenineofourtwofive\ for GW190425's primary. 
By contrast,~\PNS\ for GW190814's secondary varies by $\leq 10\%$, demonstrating robustness from its high signal-to-noise ratio and small mass ratio.
Analysis using EOS information tends to affect \PNS\ through the inferred maximum neutron star mass rather than the maximum spin. As it stands, \PNS\ remains sensitive to numerous population parameters, limiting its reliability and potentially leading to ambiguous classifications of future GW events.

\end{abstract}

\keywords{neutron star, gravitational waves, astrostatistics}
\maketitle
\section{\label{sec:intro} Introduction}
Gravitational wave (GW) astronomy has revolutionized our understanding of compact objects. 
Over the course of four observing runs, O1 \cite{collaborationGWTC1GravitationalWaveTransient2019}, O2 \cite{abbottGWTC2CompactBinary2021, collaborationGWTC21DeepExtended2022}, O3 \cite{collaborationGWTC3CompactBinary2023, collaborationPopulationMergingCompact2023}, and most recently O4a \cite{collaborationGWTC40MethodsIdentifying2025, collaborationGWTC40UpdatingGravitationalWave2025}, the LIGO-Virgo-KAGRA (LVK) collaboration has detected hundreds of compact binary mergers \cite{acerneseAdvancedVirgo2nd2015, akutsuOverviewKAGRADetector2021, collaborationAdvancedLIGO2015}. 
While binary black hole (BBH) mergers dominate the catalog, several binary neutron star (BNS) and neutron star–black hole (NSBH) systems have also been identified. 
The landmark detection of GW170817 during O2 provided the first confirmed BNS merger and catalyzed the era of multi-messenger astronomy \cite{collaborationGW170817ObservationGravitational2017}. Subsequent observing runs have added to the diversity of detected systems, including candidate BNS and NSBH mergers GW190425 \cite{collaborationGW190425ObservationCompact2020}, GW190814 \cite{fattoyevGW190814Impact262020, godziebaMaximumMassNeutron2021, mostLowerBoundMaximum2020}, GW200105 and GW200115 in O3 \cite{collaborationObservationGravitationalWaves2021} and GW230529 in O4 \cite{collaborationObservationGravitationalWaves2024}.
Notably GW190814 and GW230529 include compact objects with an inferred mass consistent with the putative \textit{lower mass gap}. 
Here we use “lower mass gap” to denote the mass range, commonly interpreted as the region between the maximum neutron star (NS) mass \cite{annalaEvidenceQuarkmatterCores2020, antoniadisMassivePulsarCompact2013a, demorestShapiroDelayMeasurement2010, fonsecaRefinedMassGeometric2021a, gandolfiMaximumMassRadius2012, kalogeraMaximumMassNeutron1996, PhysRevD.109.043052, legredImpactPSR$mathrmJ0740+6620$2021, linaresPeeringDarkSide2018, margalitConstrainingMaximumMass2017, millerRadiusPSRJ0740+66202021, oppenheimerMassiveNeutronCores1939, ozelMassesRadiiEquation2016, prakashEquationStateMaximum1988, rezzollaUsingGravitationalwaveObservations2018, rhoadesMaximumMassNeutron1974, romaniPSRJ09520607Fastest2022, shibataConstraintMaximumMass2019, tolmanStaticSolutionsEinsteins1939} and the minimum black hole (BH) mass \cite{belczynskiCygX3Galactic2013, chawlaGaiaMayDetect2022, collaborationPopulationMergingCompact2023a, farrMassDistributionStellarmass2011a, fishbachDoesMatterMatter2020, jayasingheUnicornMonoceros$3M_odot$2021, littenbergNeutronStarsBlack2015, thompsonDiscoveryCandidateBlack2019, woosleyEvolutionMassiveHelium2019, yeInferringNeutronStar2022a}.  
Its existence and precise bounds remain under active debate \cite{yeLowermassgapBlackHoles2024, xingMassgapBlackHoles2025}.
As a result, mergers in this range serve as key probes of the NS to BH transition, a region where compact object formation is still uncertain.

The idea of a lower mass gap predates gravitational wave observations and was first motivated by the absence of $3 - 5\Msun$ black holes in X-ray binaries \cite{fryerCOMPACTREMNANTMASS2012a,kreidbergMassMeasurementsBlack2012, farrMassDistributionStellarMass2011}.
A plausible explanation for the gap could be due to core-collapse supernovae, in which, following the formation of the proto-NS, the efficiency of neutrino heating plays a key role in fallback accretion. 
Sufficient neutrino heating drives a successful explosion that ejects most of the overlying material, yielding a stable NS (i.e.,rapid timescale accretion $\sim$~10ms). 
In contrast, insufficient heating leads to shock stagnation and substantial fallback (i.e., delayed timescale accretion~$\sim$~200ms), driving the remnant beyond the stability threshold and producing a BH that is typically several solar masses more massive \cite{belczynskiMISSINGBLACKHOLES2012a, fryerCOMPACTREMNANTMASS2012a, fryerEffectSupernovaConvection2022, kochanekFAILEDSUPERNOVAEEXPLAIN2014, liuFinalCompactRemnants2021, gaoFormationMassgapBlack2022, siegelInvestigatingLowerMass2023}.
Both GW and electromagnetic detections have begun to challenge the existence of a sharp gap \cite{antoniadisMassivePulsarCompact2013, belczynskiMissingBlackHoles2012, collaborationGW190814GravitationalWaves2020, demorestTwosolarmassNeutronStar2010, farahBridgingGapCategorizing2022, farrMassDistributionStellarMass2011, fonsecaRefinedMassGeometric2021, fryerCompactRemnantMass2012, kreidbergMassMeasurementsBlack2012, thompsonNoninteractingLowmassBlack2019, yeInferringNeutronStar2022, Mahapatra_2025}, with events such as GW190425, GW190814 and GW230529 consistent with component masses extending into the gap. 
These detections suggest that the lower mass gap may not be completely empty, but rather sparsely populated. 
Yet the interpretation of these events remains uncertain. 
For example, the classification of GW230529's primary may depend on the assumed spin and mass priors \cite{markinChallengingBinaryNeutron2025} or the inclusion of tidal measurements in the GW waveform (recovery template). These systematics are explored in \citet{cotturoneCharacterizingCompactObject2025}.

The NS equation of state (EOS) sets the theoretical maximum mass for a stable NS \cite{akmalEquationStateNucleon1998, douchinUnifiedEquationState2001, finchUnifiedNonparametricEquationofstate2025, friedmanHotColdNuclear1981, golombInterplayAstrophysicsNuclear2025a, lalazissisNewParametrizationLagrangian1997, landryNonparametricConstraintsNeutron2020, mohantyAstrophysicalConstraintsNeutron2024, ngInferringNeutronStar2025, wiringaEquationStateDense1988}. 
Rapid rotation can increase the support for heavy NS by 30\% \cite{13075191PromptMerger, breuMaximumMassMoment2016}. 
Collectively, these EOS-dependent effects introduce uncertainty in the maximum mass a NS can support.

In this work, we investigate how assumptions about the underlying population of compact binaries influence the classification of low mass events. 
For events like GW190425 or GW230529, where the component masses are consistent with both NSs and BHs, the inferred classification can shift significantly depending on the assumed population model. 
By explicitly studying the population-level distributions over masses and spins, we assess how these assumptions propagate into event-level classification.

Explicitly, a population model that enforces a sharp mass gap or favors anti-aligned spin between the two compact objects can suppress support for higher-mass events, even when the data alone remains ambiguous.
We systematically explore how variations in population assumptions affect the classification of compact objects as either NS or BH, \utty{noting that some aspects of the inferred population, particularly at the low-mass end, may be weakly constrained.}
We henceforth refer to this as~\PNS, with $P(m<\mmaxpop)$ representing the population-only analysis and $P(m<\mmaxeos)$ denoting the EOS-informed analyses. Here, $\mmaxpop$ and $\mmaxeos$ correspond to the boundaries taken to represent the NS-BH transition in their respective cases, with the former motivated by features in the inferred mass distribution and the latter by the maximum mass supported by the NS EOS.

The remainder of this paper is structured as follows. 
In Section~\ref{sec:methods}, we describe our methodology for hierarchical Bayesian inference, including a flexible parametric population model (\popmodel~and extensions) \cite{fishbachDoesMatterMatter2020, farahBridgingGapCategorizing2022, maliStrikingChordSpectral2024, collaborationGWTC40PopulationProperties2025}. 
We then outline the calculation of~\PNS~based on \citet{essickDiscriminatingNeutronStars2020}, including extensions to account for maximally spinning NS.
In Section~\ref{sec:mainresults}, we present our main population fits and~\PNS~results, including a comprehensive “classification matrix” (Table~\ref{tab:feature_summary}) that maps how population parameters affect classification. 
We also highlight our main results in Table~\ref{tab:fullresults} \utty{and additional information in Table~\ref{app:PCItable}}. 
These tables forms the conceptual backbone of the paper, motivating a series of focused investigations into specific event-level degeneracies.

Sections~\ref{sec:qm1m2} and \ref{sec:qXeff} explore the dominant event-level degeneracies \qmone, \qmtwo, and \qXeff. 
These sections investigate classification outcomes, particularly for low SNR events like GW230529 where degeneracies are more pronounced. 
We show that in high SNR events like GW190814, some of these degeneracies begin to break. 
In Section~\ref{sec:lookingahead}, we examine potential extensions of this work. 
\utty{Additional methodological details, including the definition of our population model and a detailed discussion of discrepancies between our GW230529 results and those reported in \citet{collaborationObservationGravitationalWaves2024}, are provided in the Appendices.}

\section{\label{sec:methods} Methodology}
\subsection{Hierarchical Bayesian Inference}
We construct a hierarchical population model and use \numevents~confidently detected compact-binary coalescences (CBCs) assigned a False-Alarm-Rate (FAR) of less than 0.25 per year by at least one search within the LVK's third gravitational wave catalog (GWTC-3) to simultaneously infer the joint mass, redshift, and spin astrophysical distributions.
We assume CBCs follow an inhomogeneous Poisson process and marginalize over the overall rate of mergers. 
Given parameters that describe the CBC merger density and other population parameters ($\Lambda$), the likelihood of observed data ($d_i$) for each event ($i$) is then 
\begin{equation}\label{eq:likelihood}
    p(\{d_i\}| \Lambda) \propto \frac{1}{\mathcal{E}^{N}}\prod_{i=1}^{N}\mathcal{Z}_{i} 
\end{equation}
where we have defined the single-event evidence $\mathcal{Z}_i$
\begin{equation}\label{eq:Zi}
    \mathcal{Z}_{i}(\Lambda) = p(d_i|\Lambda) = \int p(d_i|\theta) p(\theta|\Lambda)  d\theta
\end{equation}
and the probability of detection $\mathcal{E}$
\begin{equation}\label{eq:E}
    \mathcal{E}(\Lambda) = P(\mathbb{D}|\Lambda) = \int P(\mathbb{D}|\theta) p(\theta|\Lambda)  d\theta
\end{equation}
Within these integrals, $\theta$ represents the single-event parameters, like component masses, redshift and spins, $p(d_i|\theta)$ is the likelihood of obtaining $d_i$ given a signal described by $\theta$, and $P(\mathbb{D}|\theta)$ is the probability that a signal described by $\theta$ would be detected~marginalized over noise realizations \cite{essickDAGnabbitEnsuringConsistency2023, essickSemianalyticSensitivityEstimates2023, essickCompactBinaryCoalescence2025}.
See Refs. \cite{callisterObservedGravitationalWavePopulations2024, mandelExtractingDistributionParameters2019,mandelParameterEstimationGravitational2010, skillingNestedSampling2004, thraneIntroductionBayesianInference2019} for reviews.

Our set of \numevents~confident events from GWTC-3 were obtained by selecting events for which at least one search (either cWB \cite{dragoCoherentWaveBurstPipeline2021}, GstLAL \cite{cannonGstLALSoftwareFramework2020}, MBTA \cite{aubinMBTAPipelineDetecting2021}, and one of two PyCBC searches \cite{usmanPyCBCSearchGravitational2016}) reported a false alarm rate (FAR) $\leq 0.25/\mathrm{year}$. 
Since the updated GWTC-4.0 catalog contains no new BNS, we have not included the new events in our population fit.
However, we include GW230529 in our classification tests. 
A longer explanation can be found in Appendix~\ref{app:GWTC3vs4}.  
We use the joint O1+O2+O3 sensitivity estimates, where O1 and O2 estimates are semi-analytic and O3 estimates include real search sensitivity \cite{ligo_scientific_collaboration_and_virgo_2021_5636816}. 

Additionally, we consider events across the entire mass spectrum, including BNS, NSBH, and BBH coalescences. 
Some other analyses have focused on either just the BNS or NSBH mergers, typically applying mass cuts while not always accounting for the corresponding changes in the detection probability \(P(\mathbb{D}|\Lambda)\). 
We instead treat selection effects consistently \cite{ligo_scientific_collaboration_and_virgo_2021_5636816}. 
We find that the BBH portion of the mass and spin distributions weakly links to BNS and NSBH events and therefore has minimal effect on classification. We nevertheless use the entire CBC population to capture the full behavior within the lower mass gap.

We adopt the \popmodel~population model outlined in Equation~\ref{eq:joint mass} and Figure~\ref{fig:pop_model} to describe the mass, spin, and redshift distributions of compact binaries.
This model builds upon the framework first introduced by \citet{fishbachDoesMatterMatter2020}, with subsequent extensions presented in \citet{farahBridgingGapCategorizing2022}, \citet{maliStrikingChordSpectral2024} and finalized in GWTC-4.0 \cite{collaborationGWTC40PopulationProperties2025}. 
Given that our analysis is based on \textsc{GWTC-3}, we do not explicitly model an upper mass gap because the catalog contains too few high-mass events to meaningfully constrain such a feature \cite{collaborationPopulationMergingCompact2023b}.

For the redshift distribution (Equation~\ref{eq:p of z}), we assume mergers are uniformly distributed in comoving volume and source-frame time, with a redshift-dependent merger rate that scales as $(1 + z)^{\kappa}$.

The spin magnitude distribution (shown in Equation~\ref{eq:spinmag} and Figure~\ref{fig:popspin}) for $\chi_1, \chi_2$ is modeled using separate truncated Gaussians, distinguishing between binaries that may contain at least one NS and those composed purely of BHs (i.e., the spin magnitude depends on the component mass with a switch point $m_{\rm break,spin} = 3$\Msun). 
Similarly, the spin tilt described by $\cos\theta_1, \cos\theta_2$ is outlined by Equation~\ref{eq:spintilt}. It uses a mixture model combining a uniform distribution with a truncated Gaussian. 
As with the spin magnitudes, we adopt distinct tilt distributions for NS containing systems and for BH-only binaries to reflect differences in their likely formation channels \cite{sonNotJustWinds2024} (i.e., mass dependent distribution for $\cos\theta_i$ with a switch point $m_{\rm break,spin} = 3$\Msun).

The entire set of population hyperparameters along with the respective priors are described in Appendix~\ref{app:popmodel}.

\subsection{Classifying Compact Objects}
\citet{essickDiscriminatingNeutronStars2020} introduce a mass based framework for estimating whether a compact object is consistent with a NS along with its uncertainty. 
This was generalized to a joint mass–spin framework within the LVK's analysis of GW230529 \cite{collaborationObservationGravitationalWaves2024}. 
The method is implemented in the open-source \texttt{mmax-model-selection} package \cite{mmax-model-selection}, which builds upon \texttt{gw-distributions} \cite{gw-distributions}. 
This analysis assumes that all objects that have masses and spins consistent with a NS are NSs, which consequently results in upper limits on \PNS.

We begin by defining a region in the mass-spin parameter space where NSs exist. 
\begin{equation}\label{eq:thetagammalow}
    \Theta_{\text{NS}}(m) = \Theta(m \leq \gammalow)
\end{equation}
Here, $\Theta$ is the Heaviside step function, $m$ is the object's mass and \gammalow~is a mass scale in the population that determines the start of the lower mass gap \cite{farahBridgingGapCategorizing2022}. 
Alternatively we can compare against mass and spin limits set by the EOS. 

\begin{equation}
\begin{split}
\Theta_{\text{NS}}(m, \chi) 
&= \Theta(m \leq m_{\max}(\mathrm{EOS}, \chi)) \\
&\quad \times \Theta(\chi \leq \chi_{\max}(\mathrm{EOS}))
\end{split}
\label{eq:indicator}
\end{equation}
Above, $m_{\max}$ and $\chi_{\max}$ denote the EOS-dependent upper limits on NS mass and spin. 
To model these limits, we adopt empirical relations for the maximum mass of rotating NS taken from \citet{breuMaximumMassMoment2016}. 
\begin{equation}
\chi_{\max} = 0.5543 \, C_{\text{TOV}}^{-1/2}
\label{eq:amax}
\end{equation}
\begin{equation}
\frac{m_{\max}}{m_{\text{TOV}}} = 1 + 0.4283 \, C_{\text{TOV}} \, \chi^2 + 0.7533 \, C_{\text{TOV}}^2 \, \chi^4
\label{eq:mmax}
\end{equation}
where $C_{\text{TOV}} = G m_{\text{TOV}} / (c^2 r_{\text{TOV}})$ is the compactness of a non-rotating NS at the Tolman-Oppenheimer-Volkoff (TOV) mass limit; $m_{\text{TOV}}$ and $r_{\text{TOV}}$ are the corresponding mass and radius at that limit.

To estimate the probability that an object falls within this allowed region, we consider a joint posterior distribution over the population hyperparameters $\Lambda$, the $\EOS$, and the event-level parameters $\theta$ (which include $m$ and $\chi$) given some gravitational wave catalog $H$ which includes the data for individual events \footnote{In theory, $H$ can be any background information.}. 
The probability that the object is consistent with a NS is then:
\begin{equation}
\begin{split}
P(\text{NS}) 
&= \int d\Lambda \, d\EOS \, d\theta \, p(\Lambda, \EOS, \theta | H) \, \Theta_{\text{NS}}(\theta) \\
&= \int\, d\Lambda \, d\EOS \, p(\Lambda, \EOS | H) \int\ d\theta \, p(\theta | \Lambda, H)\,\, \Theta_{\text{NS}}(\theta) 
\end{split}
\label{eq:prob_ns}
\end{equation}
This decomposition separates the joint distribution into a population-level component and an event-level component, which is often convenient for handling separate population and event-level samples. 
For each event of interest $i$, the event-level posterior is simply the single-event posterior conditioned on its data $d_i$: $p(\theta | \Lambda, H) = p(\theta | \Lambda, d_i)$. 

\utty{Equation~\ref{eq:prob_ns} makes clear that each value of~\PNS~is obtained by marginalizing jointly over population hyperparameters, EOS uncertainty (when included), and single-event evidence.
In particular, even when the population hyperposterior is narrow, the single-event evidence can substantially contribute to variance in~\PNS, and vice versa. See 
Appendix~\ref{app:mmmsextra} for more details.}

\utty{In the EOS-informed analysis, we use a publicly available set of Gaussian Process EOS samples \cite{legred_2022_6502467}, originally presented in \citet{legredImpactPSR$mathrmJ0740+6620$2021}. They are designed to provide a flexible and model-agnostic behavior of NS matter. The resulting EOS intentionally cover a broad range of mass--radius relations and maximum NS masses consistent with current observations. 
}

\begin{figure}[h!]
\setlength{\abovecaptionskip}{2pt}
\includegraphics[width=\columnwidth]{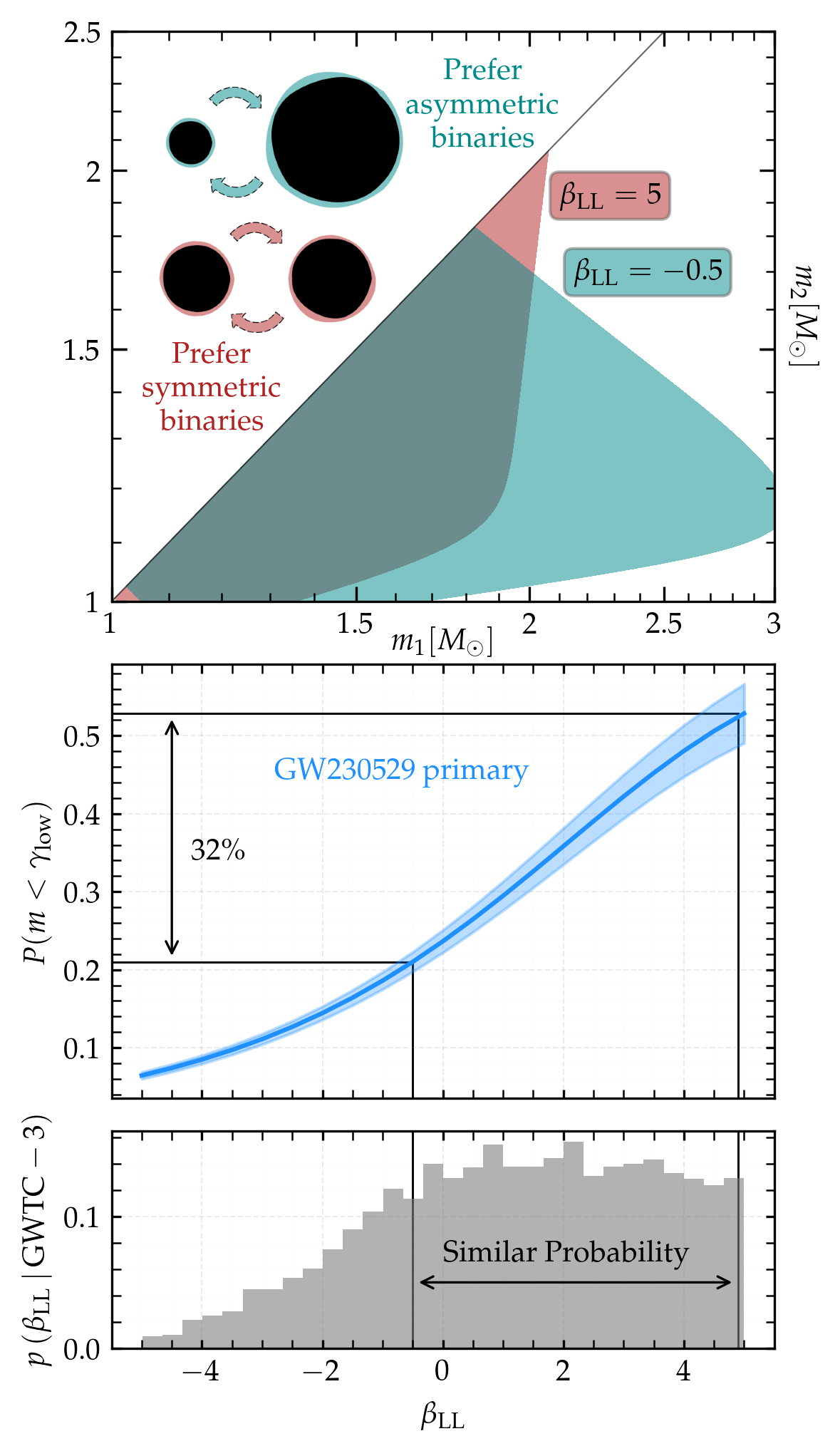}
\caption{\label{fig:betalowexample} The pairing of low mass compact objects (\pairingBNS) effect on classification of GW230529's primary. 
Here, “pairing” refers to the low-mass population hyperparameter \pairingBNS\ that governs how strongly binaries prefer near-equal component masses.
\textbf{(Top)} Joint BNS mass distribution $p(m_1,m_2|\pairingBNS,\textsc{GWTC-3})$ evaluated across pairing functions with comparable posterior support. Strong pairing ($\pairingBNS = 5$) favors near equal masses, while weak pairing ($\pairingBNS = -0.5$) skews toward more asymmetric binaries. \textbf{(Middle)} The probability of classifying GW230529's primary as a NS for various \pairingBNS; \gammalow~represents the start of the NS-BH boundary. The shaded region represents the 90\% symmetric credible interval around the mean. \PNS~varies by up to 32\% for equally likely~\pairingBNS~and between~\GWtwothreezerofivetwonineprimarybetalowNSminPNS\% -- \GWtwothreezerofivetwonineprimarybetalowNSmaxPNS\% in total. \textbf{(Bottom)} Hyperposterior $p(\pairingBNS|\rm{GWTC-}3)$. A wide range of~\pairingBNS~values have comparable probabilities. 
\pairingBNS~spans values that meaningfully affect the mass distribution; for larger \pairingBNS, the inferred $p(m_1,m_2|\pairingBNS,\textsc{GWTC-3})$ changes very little.
}
\end{figure}

Examples of~\PNS~can be found in Figure~\ref{fig:betalowexample}, which shows how \PNS\ varies across different realizations of the pairing preferences of NSs with other compact objects, encoded by the hyperparameter \pairingBNS. Here, “pairing” refers to a NS’s preference to be partnered with another compact objects of a similar mass. Larger \pairingBNS\ favors equal-mass BNS pairings, whereas smaller values favor unequal-mass systems. Current data place only weak constraints on \pairingBNS, so very different pairing behaviors have similar posterior probability. Nonetheless these hyperparameter values can drive large swings in \PNS.


\section{\label{sec:mainresults} Main results}

\begin{table*}[t]
\caption{\label{tab:feature_summary} Summary of how key population features affect the probability that a component is a NS, \PNS, under two modeling choices: population-only $P(m < \mmaxpop)$ and EOS-informed $P(m < \mmaxeos)$. Ticks (\tickonetxt–\tickthreetxt) denote the effect of a hyperparameter on classification with \tickonetxt~being a weak correlation and~\tickthreetxt~being a very strong correlation: \crosstxt\ denotes negligible relation to~\PNS. Each feature is tied to specific hyperparameters in the population model: the lower and upper edges of the NS–BH dip are controlled by~\gammalow~and~\gammahigh; the equal mass preference (“pairing”) by \pairingBNS; the spin magnitude distribution by~\muchi~(mean) and~\sigchi~(width); the spin tilt distribution by~\mutilt and~\sigtilt; and the NS mass peak with~\mupeakNS~(location) and \sigpeakNS~(width). \pairingBNS\ and \mutilt\ strongly affect~\PNS. The strong correlation with~\gammalow~in the population-only analysis is expected as this directly compares objects masses to~\gammalow~(see Sec.~\ref{sec:gammalownote}).}

\centering
\renewcommand{\arraystretch}{1.5}
\begin{tabular}{llccccccc}
\toprule
\multirow{2}{*}{\textbf{\TableSize{Param}}} 
  & \multirow{2}{*}{\textbf{\TableSize{Physical Feature}}}
  & \multirow{2}{*}{\textbf{\TableSize{Refs}}}
  & \multicolumn{2}{c}{\textbf{\TableSize{Low SNR}}} 
  & \multicolumn{2}{c}{\textbf{\TableSize{Asymmetric Masses}}} \\ 
  \cmidrule(r){4-5} \cmidrule(r){6-7}
  & & 
  & \TableSize{\poplabel} & \TableSize{\eoslabel} 
  & \TableSize{\poplabel} & \TableSize{\eoslabel} \\
\midrule
\gammalow             & \TableSize{Lower edge of the NS–BH Dip}        & (Sec~\ref{sec:gammalownote}) & \tickthree & \cross     & \tickthree     & \cross     \\
\gammahigh            & \TableSize{Upper edge of the NS–BH Dip}        & (Sec~\ref{sec:gammahigh}) & \cross     & \cross     & \cross & \cross     \\
\pairingBNS           & \TableSize{Pairing between BNS}                & (Sec~\ref{sec:qm1m2_pairing}) & \tickthree & \tickthree   & \tickone   & \tickone    \\
\muchi, \sigchi       & \TableSize{Spin Magnitude distribution of NS}  & (Sec~\ref{sec:qXeff_mag}) & \ticktwo   & \ticktwo     & \cross & \cross   \\
\mutilt, \sigtilt     & \TableSize{Spin Tilt distribution of NS}       & (Sec~\ref{sec:qXeff_tilt}) & \tickthree   & \ticktwo   & \tickone     & \tickone   \\
\mupeakNS, \sigpeakNS & \TableSize{Peak in the NS mass distribution}   & (Sec~\ref{sec:qm1m2_sigpeakNS}) & \ticktwo & \tickone & \cross & \cross     \\
\bottomrule
\end{tabular}
\end{table*}
In this section, we proceed in three steps. First, we summarize the population model. Next, we map how key hyperparameters influence \PNS, with the key features highlighted in Table~\ref{tab:feature_summary}. We then quantify the resulting shifts in \PNS\ for each hyperparameter across events, reported in Table~\ref{tab:fullresults}. \utty{Each row corresponds to varying a single hyperparameter, and each table entry reports the resulting range of~\PNS\ for that event. Columns labeled ``Pop'' and ``EOS'' show results obtained using the population-only and EOS-informed classification schemes, respectively. Larger ranges indicate greater sensitivity of~\PNS~to the corresponding population feature. }

We do not consider GW170817 in detail, as its high SNR and relatively low component masses make its classification as a BNS essentially certain across models. \utty{
We include GW190425 and GW230529 as they are low SNR systems with component masses near the lower mass gap. We also include GW190814 to provide a contrasting high SNR, asymmetric-mass system with one component near the gap. 
Although we also analysed other NSBH candidates, such as GW200105 and GW200115, we do not include them here, as they are marginal and do not qualitatively alter our conclusions.
}

\subsection{Inferred population model}\label{sec:popmodels}
We use \textsc{FullPop4.0} population model, with an additional NS peak, a modified spin prescription, and the standard redshift-evolution model (see Appendix~\ref{app:popmodel} for details). 
Figure~\ref{fig:popmass} shows the inferred latent mass distribution across the entire mass spectrum \cite{collaborationGWTC40PopulationProperties2025}.
The curve rises steeply at 1-2\Msun, then dips sharply between 3-5\Msun~before recovering the familiar BBH regime. 
The inclusion of an explicit NS-peak does not greatly alter the overall shape of the mass distribution and remains consistent with expectation from existing population studies \cite{maliStrikingChordSpectral2024,collaborationGWTC40PopulationProperties2025}.
\utty{The precise behavior of the distribution at the low-mass edge of this dip is only weakly constrained by data and remains sensitive to prior assumptions, an issue we return to in Section~\ref{sec:gammalownote}.}

\begin{figure}
    \centering    \includegraphics[width=\columnwidth]{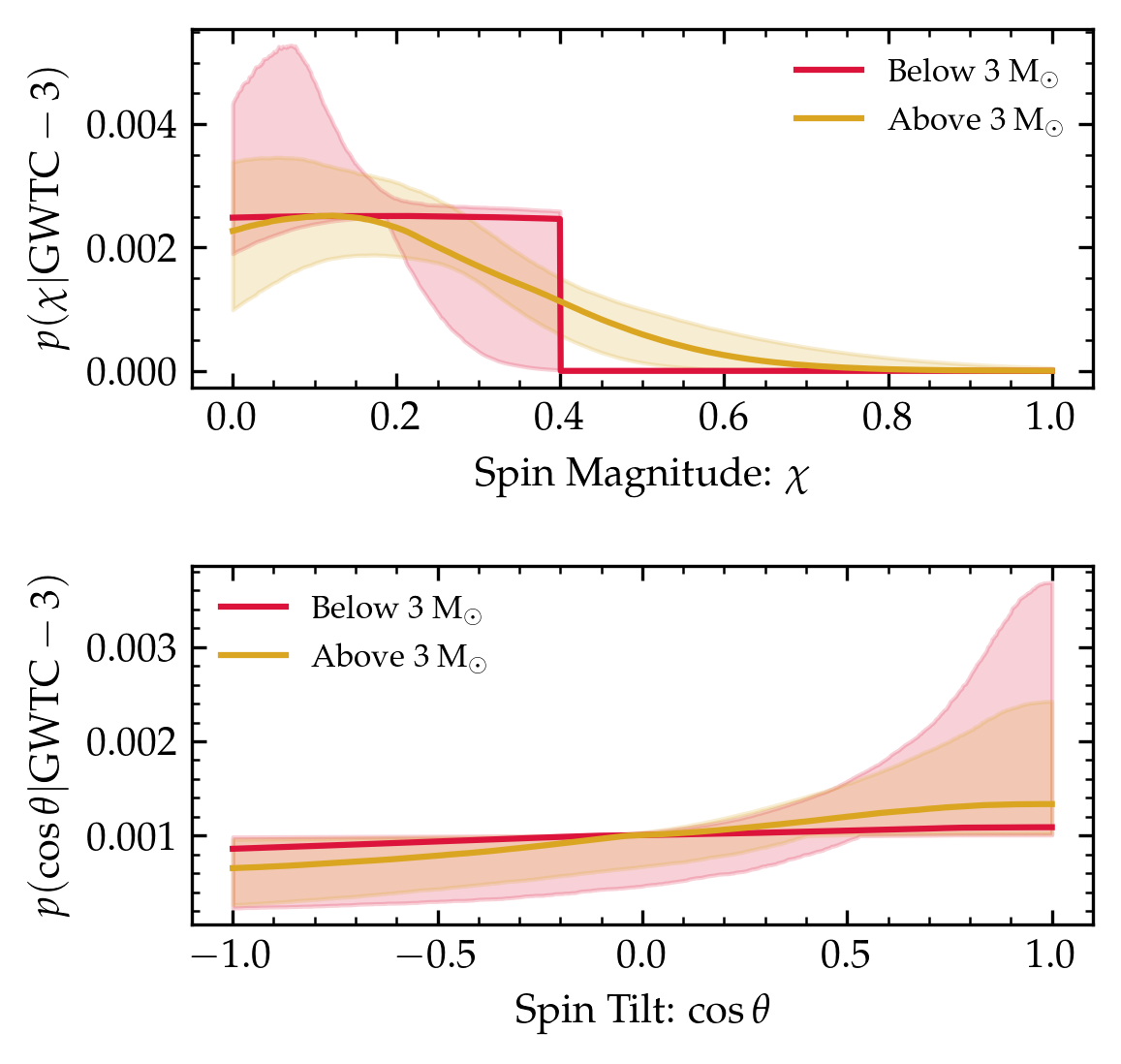}
    \caption{Population spin distributions inferred from GWTC-3. (\textbf{Top}) Spin magnitude distributions for components below and above $3\,M_\odot$. (\textbf{Bottom}) Corresponding spin tilt ($\cos\theta$) distributions. The solid curve shows the mean, while the shaded band denotes the 90\% symmetric credible interval. 
}
    \label{fig:popspin}
\end{figure}

Figure~\ref{fig:popspin} presents the inferred spin distributions split into two mass regimes to capture potentially distinct physics above and below $3\,M_\odot$. \utty{We assume that the component spins are conditionally independent. While not required in principle, relaxing this assumption would introduce additional population uncertainty.} The upper panel shows the spin magnitude distribution, $p(\chi\,|\,\mathrm{GWTC}\text{-}3)$, while the lower panel shows the tilt distribution, $p(\cos\theta\,|\,\mathrm{GWTC}\text{-}3)$. For objects below $3\,M_\odot$ (red), the spin magnitude distribution is confined to $\chi \leq 0.4$, a limit set by the O3 injection coverage~\cite{essickCompactBinaryCoalescence2025}. Within this range, the distribution peaks at low spin values, consistent with expectations for NSs. For higher-mass objects (yellow), the distribution has a longer tail extending to moderate spins, indicative of the broader spin support in the BH population. In the tilt distribution (bottom panel), both mass regimes show broadly overlapping uncertainty bands and a mild preference for aligned spins. This mass-dependent break in the spin distributions allows the analysis to isolate spin-dependent physics for low mass objects.

We do not show our redshift model as it is identical to many other population analysis \cite{collaborationGWTC40PopulationProperties2025,ligo_scientific_collaboration_and_virgo_2021_5636816, callisterObservedGravitationalWavePopulations2024}. Refer to Appendix~\ref{app:popmodel} for a detailed outline of all models. Additionally, we include the full joint correlations between parameters in the corner plot shown in Figure~\ref{fig:corner}.  

\begin{figure}
    \centering
\includegraphics[width=\columnwidth]{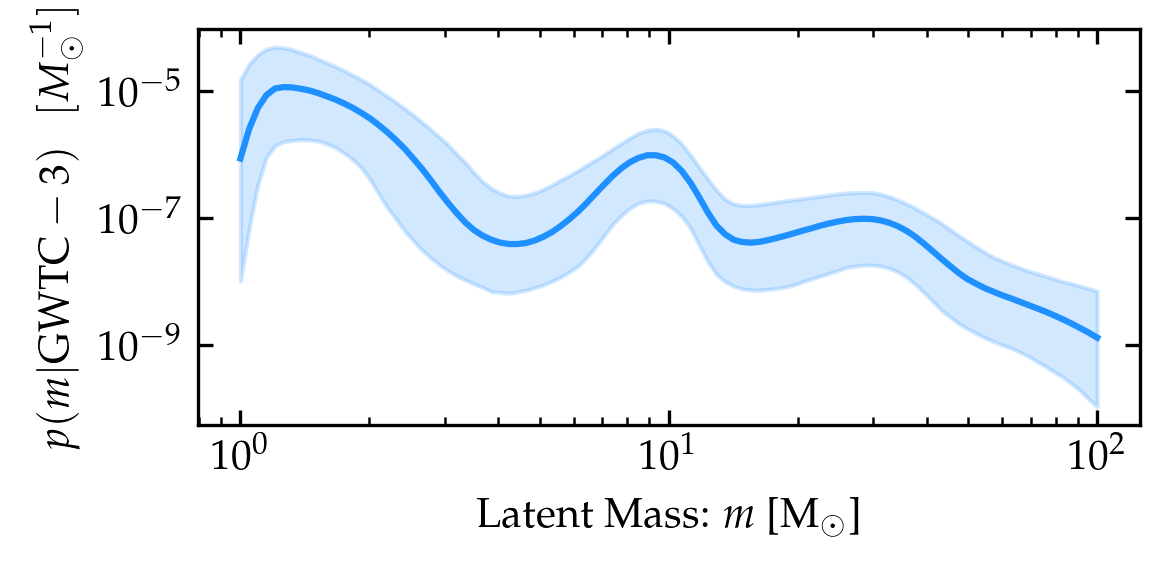}
    \caption{Inferred latent mass distribution $p(m)$ from GWTC-3.0. The solid curve shows the mean, while the shaded band denotes the 90\% symmetric credible interval. A clear dip appears across the putative NS–BH gap, with BBH peaks near $m \sim 10M_\odot$ and $m \sim 30M_\odot$.}
    \label{fig:popmass}
\end{figure}

\subsection{Inferred neutron star probabilities~\PNS}
\label{app:resultstable}
We now move from population-level structure to event-level consequences. 
For each event, \PNS\ is evaluated by marginalizing over hyperposterior samples from our population fit (population-only: $P(m<\mmaxpop)$; EOS-informed: $P(m<\mmaxeos)$). 
To compute our results (summarized in Table~\ref{tab:fullresults} \utty{and Table~\ref{app:PCItable}}), we artificially fix a single hyperparameter at a trial value while drawing the remaining hyperparameters from their hyperposterior. 
This isolates the response of \PNS\ to that specific feature (e.g., pairing preferences, spin magnitudes/tilts etc). 
We then iterate this over the prior range of the trial hyperparameter to generate an illustrative plot. 
Although this procedure does not reflect the fully self-consistent joint hyperposterior (because one parameter is held fixed without conditioning the posterior distribution over the rest of the hyperparameters), it is informative about the magnitude and direction of the changes to~\PNS. 
We repeat this method across the range of hyperparameters outlined in Table~\ref{tab:fullresults}.

Uncertainties on \PNS\ arise from the finite Monte Carlo variance from a limited number of hyperposterior draws and event-level posterior samples.

Across the two classification schemes, we find that changes in the population model have broadly comparable impacts on \PNS\ for population-only ($P(m<\mmaxpop)$) and EOS-informed ($P(m<\mmaxeos)$) analyses. We next examine joint structure in the $\q$--$\mone$, $\q$--$\mtwo$, and $\q$--$\Xeff$ planes.

\begin{table*}[t]
\centering
\caption{\label{tab:hyper_summary}Summary of classification outcomes for key hyperparameters across GW190425's primary, GW190814's secondary and GW230529's primary. We compare both the population-only (Equation~\ref{eq:thetagammalow}) and EOS-informed (Equation~\ref{eq:indicator}) analyses. We show the range of~\PNS~spanned when varying each parameter, with the bottom row reporting the overall maximum change. 
\utty{For each row, we fix
the listed hyperparameter to a grid of trial values while drawing the remaining
hyperparameters from their hyperposterior (i.e. not conditioned on the fixed value). We then compute~\PNS~(Equation~\ref{eq:prob_ns}).}
\utty{We note that constraints on \gammalow\ are largely prior dominated; see Section~\ref{sec:gammalownote} for further discussion. The prior ranges and posterior credible intervals for each parameter are included in Table~\ref{app:PCItable}}. 
In order to compute the largest change, we used the most extreme settings (see Section~\ref{sec:gowild}).}
\label{tab:fullresults}
\setlength{\tabcolsep}{6pt}
\renewcommand{\arraystretch}{1.25}
\begin{tabular}{l l cc cc cc}
\toprule
 &  & \multicolumn{2}{c}{\textbf{GW190425} primary} & \multicolumn{2}{c}{\textbf{GW190814} secondary} & \multicolumn{2}{c}{\textbf{GW230529} primary} \\
\cmidrule(lr){3-4}\cmidrule(lr){5-6}\cmidrule(lr){7-8}
\textbf{Param} & \textbf{Description} & \textbf{Pop (\%)} & \textbf{\eoslabel (\%)} & \textbf{Pop (\%)} & \textbf{\eoslabel (\%)} & \textbf{Pop (\%)} & \textbf{\eoslabel (\%)} \\
\midrule
$\sigpeakNS$ & Width of peak in NS mass dist &
$\GWoneninezerofourtwofiveprimarysigpeakNSminPNS$ -- $\GWoneninezerofourtwofiveprimarysigpeakNSmaxPNS$ &
$\GWoneninezerofourtwofiveprimarysigpeakNSEOSminPNS$ -- $\GWoneninezerofourtwofiveprimarysigpeakNSEOSmaxPNS$ &
$\GWoneninezeroeightonefoursecondarysigpeakNSminPNS$ -- $\GWoneninezeroeightonefoursecondarysigpeakNSmaxPNS$ &
$\GWoneninezeroeightonefoursecondarysigpeakNSEOSminPNS$ -- $\GWoneninezeroeightonefoursecondarysigpeakNSEOSmaxPNS$ &
$\GWtwothreezerofivetwonineprimarysigpeakNSminPNS$ -- $\GWtwothreezerofivetwonineprimarysigpeakNSmaxPNS$ &
$\GWtwothreezerofivetwonineprimarysigpeakNSEOSminPNS$ -- $\GWtwothreezerofivetwonineprimarysigpeakNSEOSmaxPNS$
\\[2pt]
\pairingBNS & Pairing between BNS &
$\GWoneninezerofourtwofiveprimarybetalowNSminPNS$ -- $\GWoneninezerofourtwofiveprimarybetalowNSmaxPNS$ &
$\GWoneninezerofourtwofiveprimarybetalowNSEOSminPNS$ -- $\GWoneninezerofourtwofiveprimarybetalowNSEOSmaxPNS$ &
$\GWoneninezeroeightonefoursecondarybetalowNSminPNS$ -- $\GWoneninezeroeightonefoursecondarybetalowNSmaxPNS$ &
$\GWoneninezeroeightonefoursecondarybetalowNSEOSminPNS$ -- $\GWoneninezeroeightonefoursecondarybetalowNSEOSmaxPNS$ &
$\GWtwothreezerofivetwonineprimarybetalowNSminPNS$ -- $\GWtwothreezerofivetwonineprimarybetalowNSmaxPNS$ &
$\GWtwothreezerofivetwonineprimarybetalowNSEOSminPNS$ -- $\GWtwothreezerofivetwonineprimarybetalowNSEOSmaxPNS$
\\[2pt]
\mutilt & Spin tilt distribution of NS &
$\GWoneninezerofourtwofiveprimarymucostiltminPNS$ -- $\GWoneninezerofourtwofiveprimarymucostiltmaxPNS$ &
$\GWoneninezerofourtwofiveprimarymucostiltEOSminPNS$ -- $\GWoneninezerofourtwofiveprimarymucostiltEOSmaxPNS$ &
$\GWoneninezeroeightonefoursecondarymucostiltminPNS$ -- $\GWoneninezeroeightonefoursecondarymucostiltmaxPNS$ &
$\GWoneninezeroeightonefoursecondarymucostiltEOSminPNS$ -- $\GWoneninezeroeightonefoursecondarymucostiltEOSmaxPNS$ &
$\GWtwothreezerofivetwonineprimarymucostiltminPNS$ -- $\GWtwothreezerofivetwonineprimarymucostiltmaxPNS$ &
$\GWtwothreezerofivetwonineprimarymucostiltEOSminPNS$ -- $\GWtwothreezerofivetwonineprimarymucostiltEOSmaxPNS$
\\[2pt]
\mupeakNS & Mean of peak inc NS mass dist &
$\GWoneninezerofourtwofiveprimarymupeakNSminPNS$ -- $\GWoneninezerofourtwofiveprimarymupeakNSmaxPNS$ &
$\GWoneninezerofourtwofiveprimarymupeakNSEOSminPNS$ -- $\GWoneninezerofourtwofiveprimarymupeakNSEOSmaxPNS$ &
$\GWoneninezeroeightonefoursecondarymupeakNSminPNS$ -- $\GWoneninezeroeightonefoursecondarymupeakNSmaxPNS$ &
$\GWoneninezeroeightonefoursecondarymupeakNSEOSminPNS$ -- $\GWoneninezeroeightonefoursecondarymupeakNSEOSmaxPNS$ &
$\GWtwothreezerofivetwonineprimarymupeakNSminPNS$ -- $\GWtwothreezerofivetwonineprimarymupeakNSmaxPNS$ &
$\GWtwothreezerofivetwonineprimarymupeakNSEOSminPNS$ -- $\GWtwothreezerofivetwonineprimarymupeakNSEOSmaxPNS$
\\[2pt]
\gammahigh & Upper edge of the NS--BH dip &
$\GWoneninezerofourtwofiveprimarygammahighminPNS$ -- $\GWoneninezerofourtwofiveprimarygammahighmaxPNS$ &
$\GWoneninezerofourtwofiveprimarygammahighEOSminPNS$ -- $\GWoneninezerofourtwofiveprimarygammahighEOSmaxPNS$ &
$\GWoneninezeroeightonefoursecondarygammahighminPNS$ -- $\GWoneninezeroeightonefoursecondarygammahighmaxPNS$ &
$\GWoneninezeroeightonefoursecondarygammahighEOSminPNS$ -- $\GWoneninezeroeightonefoursecondarygammahighEOSmaxPNS$ &
$\GWtwothreezerofivetwonineprimarygammahighminPNS$ -- $\GWtwothreezerofivetwonineprimarygammahighmaxPNS$ &
$\GWtwothreezerofivetwonineprimarygammahighEOSminPNS$ -- $\GWtwothreezerofivetwonineprimarygammahighEOSmaxPNS$
\\
$\sigchi$ & Spin magnitude width of NS &
$\GWoneninezerofourtwofiveprimarysigmachiminPNS$ -- $\GWoneninezerofourtwofiveprimarysigmachimaxPNS$ &
$\GWoneninezerofourtwofiveprimarysigmachiEOSminPNS$ -- $\GWoneninezerofourtwofiveprimarysigmachiEOSmaxPNS$ &
$\GWoneninezeroeightonefoursecondarysigmachiminPNS$ -- $\GWoneninezeroeightonefoursecondarysigmachimaxPNS$ &
$\GWoneninezeroeightonefoursecondarysigmachiEOSminPNS$ -- $\GWoneninezeroeightonefoursecondarysigmachiEOSmaxPNS$ &
$\GWtwothreezerofivetwonineprimarysigmachiminPNS$ -- $\GWtwothreezerofivetwonineprimarysigmachimaxPNS$ &
$\GWtwothreezerofivetwonineprimarysigmachiEOSminPNS$ -- $\GWtwothreezerofivetwonineprimarysigmachiEOSmaxPNS$
\\[2pt]
$\muchi$ & Spin magnitude mean of NS &
$\GWoneninezerofourtwofiveprimarymuchioneminPNS$ -- $\GWoneninezerofourtwofiveprimarymuchionemaxPNS$ &
$\GWoneninezerofourtwofiveprimarymuchioneEOSminPNS$ -- $\GWoneninezerofourtwofiveprimarymuchioneEOSmaxPNS$ &
$\GWoneninezeroeightonefoursecondarymuchioneminPNS$ -- $\GWoneninezeroeightonefoursecondarymuchionemaxPNS$ &
$\GWoneninezeroeightonefoursecondarymuchioneEOSminPNS$ -- $\GWoneninezeroeightonefoursecondarymuchioneEOSmaxPNS$ &
$\GWtwothreezerofivetwonineprimarymuchioneminPNS$ -- $\GWtwothreezerofivetwonineprimarymuchionemaxPNS$ &
$\GWtwothreezerofivetwonineprimarymuchioneEOSminPNS$ -- $\GWtwothreezerofivetwonineprimarymuchioneEOSmaxPNS$
\\
\midrule
\textbf{\boldmath$\mutilt$ \& $\pairingBNS$} & Spin tilt and BNS pairing &
$\GWoneninezerofourtwofiveonebetamutiltcombinedmin$ -- $\GWoneninezerofourtwofiveonebetamutiltcombinedmax$ &
$\GWoneninezerofourtwofiveonebetamutiltcombinedEOSmin$ -- $\GWoneninezerofourtwofiveonebetamutiltcombinedEOSmax$ &
$\GWoneninezeroeightonefourtwobetamutiltcombinedmin$ -- $\GWoneninezeroeightonefourtwobetamutiltcombinedmax$ &
$\GWoneninezeroeightonefourtwobetamutiltcombinedEOSmin$ -- $13$\ &
$\GWtwothreezerofivetwonineonebetamutiltcombinedmin$ -- $\GWtwothreezerofivetwonineonebetamutiltcombinedmax$ &
$\GWtwothreezerofivetwonineonebetamutiltcombinedEOSmin$ -- $\GWtwothreezerofivetwonineonebetamutiltcombinedEOSmax$
\\
\bottomrule
\end{tabular}
\end{table*}

\section{\label{sec:qm1m2} The $\bm{\q~\leftrightarrow~\mone, \mtwo}$~plane}

\begin{figure*}[t]
\includegraphics[width=1\textwidth]{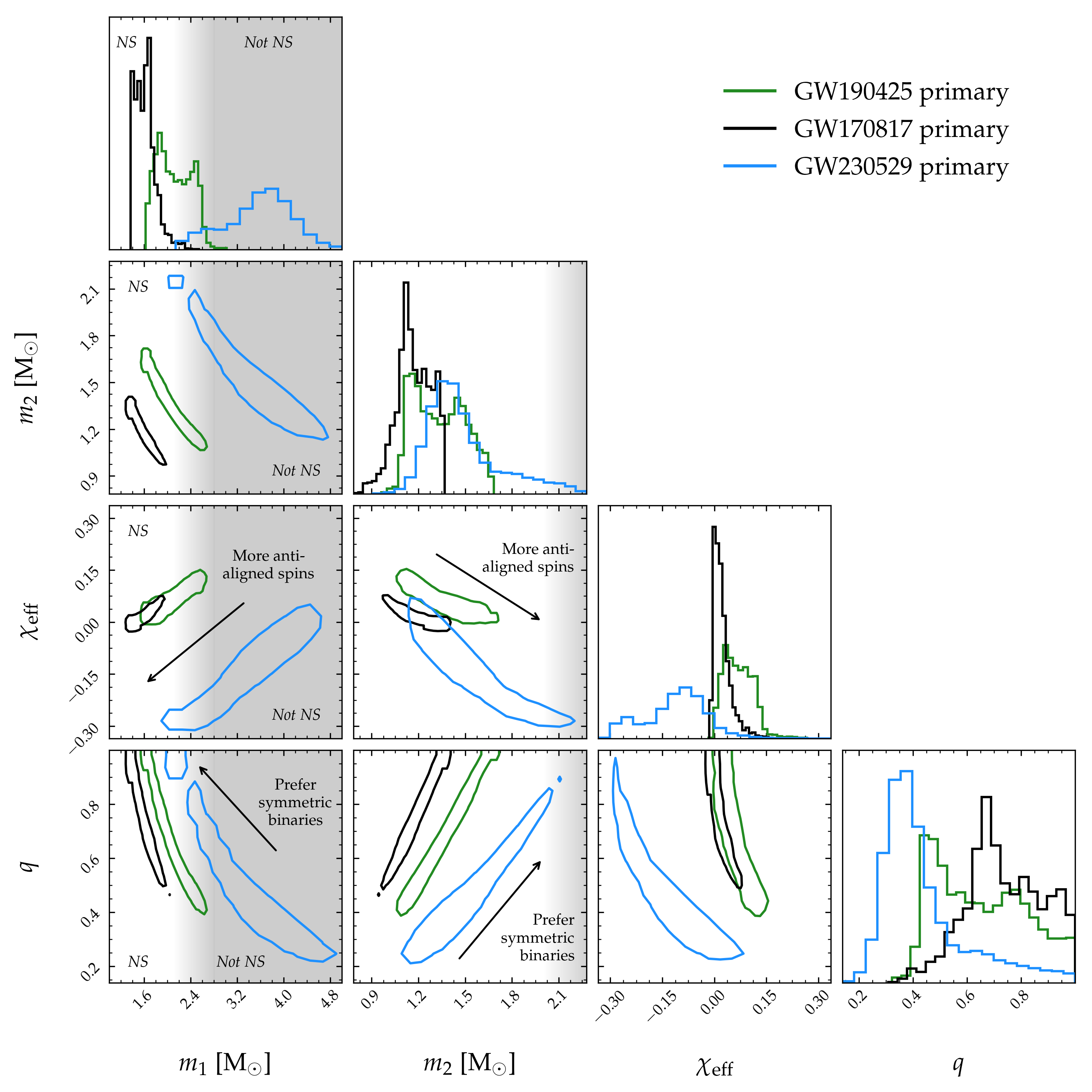}
\caption{\label{fig:correlationscartoon} Using default priors (uniform component masses in the detector frame and spins that are uniform in magnitude and isotropic in orientation), we show the inferred parameters for three events, GW170817 (red), GW190425 (green), and GW230529 (blue). Off-diagonal panels show 90\% symmetric credible intervals. Gray shading indicates regions that are incompatible with NSs (masses above the NS cutoff, \mmaxpop, or \mmaxeos~in our case) and the associated uncertainty band. Degeneracies appear between~\q, \Xeff,~\mone, and~\mtwo. Overall, the \qXeff,~\qmone, and~\qmtwo~couplings control whether the probability crosses into the ``NS'' or ``Not NS'' demarcations.}
\end{figure*}

We organize this section around correlations in two planes:~\qmone~and \qmtwo. 
This correlation arises because, for a fixed chirp mass~$\mathcal{M}_c$ (which is tightly constrained by the waveform phase evolution), changes in~$q$ require compensating shifts in~$\mone$ and~$\mtwo$ to preserve~$\mathcal{M}_c$.
Figure~\ref{fig:correlationscartoon} shows the~\qmone~and~\qmtwo~correlations using the default prior distributions used to estimate the source parameters of GW170817, GW190425 and GW230529 \cite{collaborationGW170817ObservationGravitational2017, collaborationGWTC40PopulationProperties2025, collaborationGWTC40UpdatingGravitationalWave2025}.
\utty{For GW230529, differences relative to \citet{collaborationObservationGravitationalWaves2024} arise from implementation details discussed explicitly in Appendix~\ref{app:gw230529}.}
We refer back to these corner plots throughout our discussion. 
Motion toward more symmetric mass ratios ($q \!\to\! 1$) always drives $\mone$ downward and $\mtwo$ upward. 
These directions do not depend on SNR. However, at low SNR, broad single-event posteriors allow stronger influence from priors, whereas at high SNR the single event posteriors are much narrower, limiting the effect of the $\q~\leftrightarrow~\mone$, $\mtwo$ correlations.
Below, we describe how specific hyperparameters impact~\PNS.

\subsection{The \gammalow~population case is prior dominated {\label{sec:gammalownote}}} 
Varying the lower edge of the NS-BH dip, and then counting the number of reweighted samples below it produces a 0--100\% lever on classification. By construction, any increase (decrease) in~\gammalow~includes (excludes) a portion of the reweighted single-event posterior. 
As a result, this effect does not reveal a genuine correlation in the~\q--\mone,\mtwo~degeneracy. 
For this reason, although~\gammalow~can strongly change~$P(m < \mmaxpop)$, we do not interpret it as a physical effect for our analysis, and we do not emphasize it further.

In the EOS-informed analyses, variations in~\gammalow~show no discernible correlation with $P(m < \mmaxeos)$, although the nominal values differ from the population-only results. 
This is expected, as the EOS-informed classification employs a different boundary for the NS-BH transition. Overall, $P(m < \mmaxeos)$\ varies by at most $4\%$ across events (GW230529: \GWtwothreezerofivetwonineprimarygammalowEOSminPNS\% -- \GWtwothreezerofivetwonineprimarygammalowEOSmaxPNS\%, GW190425: \GWoneninezerofourtwofiveprimarygammalowEOSminPNS\% -- \GWoneninezerofourtwofiveprimarygammalowEOSmaxPNS\% and GW190814: \GWoneninezeroeightonefoursecondarygammalowEOSminPNS~-- \GWoneninezeroeightonefoursecondarygammalowEOSmaxPNS\%, among others). 
This limited change may seem counterintuitive, given that the location of the NS-BH boundary should, in principle, influence classification. 
However,~\gammalow\ does not strongly alter the inferred shape of the mass distribution. 
The inferred ``notch amplitude'' is small ($A \sim 0.1$), so the model doesn't carve a deep low mass gap. 
Instead the mass distribution is dominated by the peaks (see Appendix~\ref{app:popmodel} for a definition of the notch amplitude). 
By contrast, \citet{farahBridgingGapCategorizing2022} infer a larger notch amplitude, yielding a deeper gap in the mass distribution. 
However, their model did not include peaks and adopts a fixed mass-independent spin distribution with a redshift distribution that did not evolve with the star formation rate.
A version of this model was used in the GW230529 discovery paper \cite{collaborationObservationGravitationalWaves2024}. 
Further discussion about the gap depth can be found in Section 4.3 of the LVK GWTC-4.0 populations paper \cite{collaborationGWTC40PopulationProperties2025}.
As a result of this, the EOS-informed classification is insensitive to variations in~\gammalow. 

\subsection{The upper edge of the NS-BH gap~minimally affects~\PNS \label{sec:gammahigh}}
Across all events considered, varying \gammahigh~produces no significant change in~\PNS. 
Consequently, we do not include a dedicated~\gammahigh~figure. 
Furthermore, since the inferred notch amplitude is small, shifting~\gammahigh~minimally changes the mass distribution.

In principle, increasing the upper edge (larger \gammahigh) could push support to higher $m_1$ and thereby pull $m_2$ lower, which might alter \PNS. 
In practice, however, our catalog lacks the specific event needed for this to occur. 
Without this ``\mone~near \gammahigh~to \mtwo~near \gammalow'' configuration, the indirect pathway for \gammahigh~to modify \PNS\ is limited.

\subsection{Pairing (preferring equal mass binaries) pushes reweighted posteriors up the \bm{$\qmone$}~plane. \label{sec:qm1m2_pairing}}
\begin{figure}[h!]
\includegraphics[width=\gridscale\columnwidth]{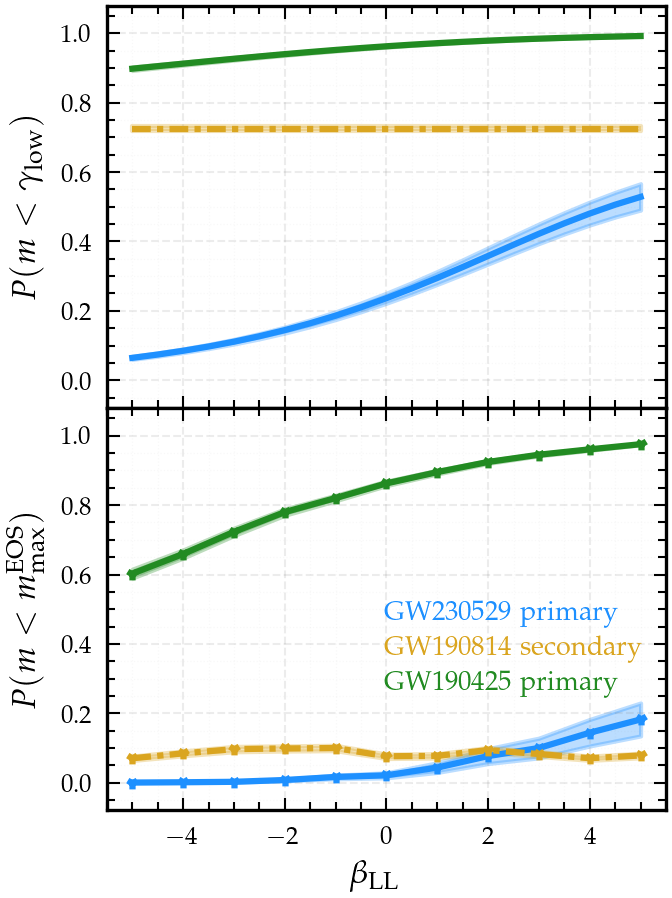}
\caption{\label{fig:beta1NS} 
The BNS pairing function's (\pairingBNS) effect on classification~\PNS.~\topbottompopeosmessage~Increasing~\pairingBNS\ strengthens equal mass pairing ($\q \rightarrow 1$). This is most apparent in GW230529's primary (top blue, shifting by $\GWtwothreezerofivetwonineprimarybetalowNSdeltapercent$\%) and GW190425 $m_1$ EOS case (bottom green, shifting by $\GWoneninezerofourtwofiveprimarybetalowNSEOSdeltapercent$\%). \shededregionmessage}
\end{figure}

The preference which low-mass objects have in merging with companions of equal mass significantly increases~\PNS~under both population-only and EOS-informed analyses. The pairing hyperparameter~\pairingBNS~acts directly on \q~through the pairing function shown in Equation~\ref{eq:pairing3},  Figures~\ref{fig:correlationscartoon} and~\ref{fig:beta1NS} . 
Increasing \pairingBNS~redistributes posterior weight toward $q\sim1$. 
In the \qmone~plane, this slides the probability down to lower \mone, and, in the~\qmtwo~plane, it pulls support toward larger~\mtwo. 
When the secondary is confidently in the NS range (typical for many ambiguous BNS and NSBH), the preference for equal-mass mergers drags the primary below the NS boundary, increasing~\PNS~for the primary. 
For GW230529, explicitly varying the pairing function can vary~\PNS\ by up to \GWtwothreezerofivetwonineprimarybetalowNSdeltapercent\%. 
Similar behavior can be observed in the EOS-informed GW190425 results that span \GWoneninezerofourtwofiveprimarybetalowNSEOSminPNS\% to \GWoneninezerofourtwofiveprimarybetalowNSEOSmaxPNS\%. 
This behavior is evident in Figures~\ref{fig:betalowexample}~and~\ref{fig:beta1NS}, where larger \pairingBNS~compresses the credible region to $q\sim1$.

While stronger pairing generally increases~\PNS~for BNS primaries, the same shift often decreases \PNS~for NSBH secondaries. 
This is a geometric consequence of moving along the~\mone--\mtwo~plane. Conversely, for high SNR asymmetric systems (e.g. GW190814),  the likelihood anchors the single-event posterior around low \q. 
Thus, the same increase in pairing will not shift the reweighted \mtwo~posterior. 
As a result, for GW190814, there is little to no change in~\PNS.  
In short,~\pairingBNS~has the largest effect on~\PNS\ when the detected event is low SNR. It also impacts $m_1$ and $m_2$ in opposite directions.

\subsection{A sharp peak in the NS mass distribution 
suppresses support for heavy NS\label{sec:qm1m2_sigpeakNS}}
The width of the NS mass peak, controlled by~\sigpeakNS, directly influences the inferred probability~\PNS. This parameter determines how tightly the population of NS is clustered around $1.4\,\Msun$. Variations in~\sigpeakNS\ can either increase or decrease~\PNS. To capture this non-monotonic behavior, we consider three representative cases for~\sigpeakNS: narrow, intermediate, and broad, corresponding to increasingly wider NS mass distributions.

\begin{figure}[h!]
\includegraphics[width=\gridscale\columnwidth]{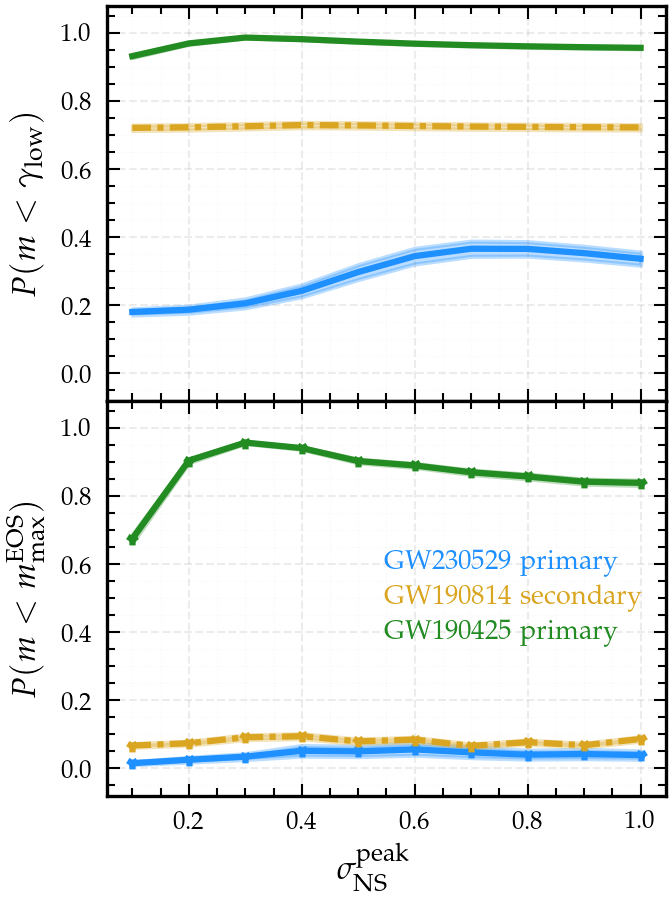}
\caption{\label{fig:sigpeakNS} Effect of the NS peak width, \sigpeakNS, on~\PNS.~\topbottompopeosmessage~Extremely narrow peaks suppress heavy NS support. Refer to Sec~\ref{sec:qm1m2_sigpeakNS} for a detailed explanation of the non-monotonic behavior in GW190425's primary for $P(m<\mmaxeos)$.~\shededregionmessage}
\end{figure}

\shortpara{Extremely narrow peak \utty{($\sigpeakNS < 0.2$)}}.
When \sigpeakNS\ is very small, the peak is sharply concentrated near $1.4\,\Msun$,~\PNS~for GW230529's primary and GW190425's primary are relatively lower.
In the \qmone~plane (Figure~\ref{fig:correlationscartoon}), this suppresses support at very low~\mone\ and collapses the credible regions toward low~\q. 
In the \qmtwo~plane, contours tighten toward lower~\mtwo. 
As shown in Figure~\ref{fig:sigpeakNS}, this region has a comparatively low \PNS~for GW190425's primary. 
A detailed treatment of the~\mone-\mtwo~relation for this event appears in \citet{foleyUpdatedParameterEstimates2020}.

\shortpara{Intermediate width peak~\utty{($0.2<\sigpeakNS<0.4$)}.}
As~\sigpeakNS is increased to the point that the mass distribution begins to influence the NS-BH boundary region ($\sim 2-3\Msun$), the broadening NS peak pulls~\mone~downward toward 1.4\Msun. 
This shifts support away from the boundary and toward a confident NS, raising~\PNS. 

\shortpara{Very broad peak \utty{($\sigpeakNS > 0.4$, quasi-uniform)}.}
If~\sigpeakNS\ is so large that the peak approaches a near uniform distribution, the probability leaks into masses beyond~\mmaxpop~(and \mmaxeos), which lowers~\PNS. 
Given the current state of observations, such an extremely wide NS mass distribution is disfavored. 

Although the NS mass peak is strongly concentrated near 1.4\Msun~\cite{kiziltanNeutronStarMass2013} and large shifts are unlikely, we vary \mupeakNS\ for completeness. The variability is weaker than \pairingBNS~and~\sigpeakNS, and sizable shifts are astrophysically disfavored, so we do not pursue it further.

\section{\label{sec:qXeff} The $\bm{\qXeff}$ plane} 
\begin{figure}[h!]
\includegraphics[width=\gridscale\columnwidth]{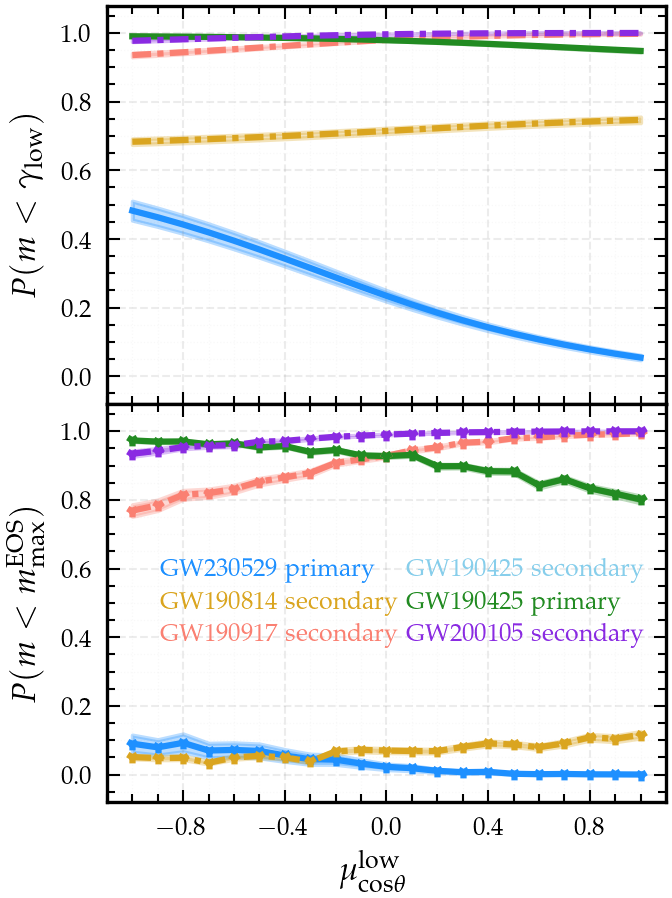}
\caption{\label{fig:mucostilt} Effect of the NS spin tilt mean~\mutilt\ on~\PNS.~\topbottompopeosmessage~Notably, GW230529 (top blue) decreases by $\GWtwothreezerofivetwonineprimarymucostiltdeltapercent$\%.~\shededregionmessage}
\end{figure}

\begin{figure}[h!]
\includegraphics[width=\gridscale\columnwidth]{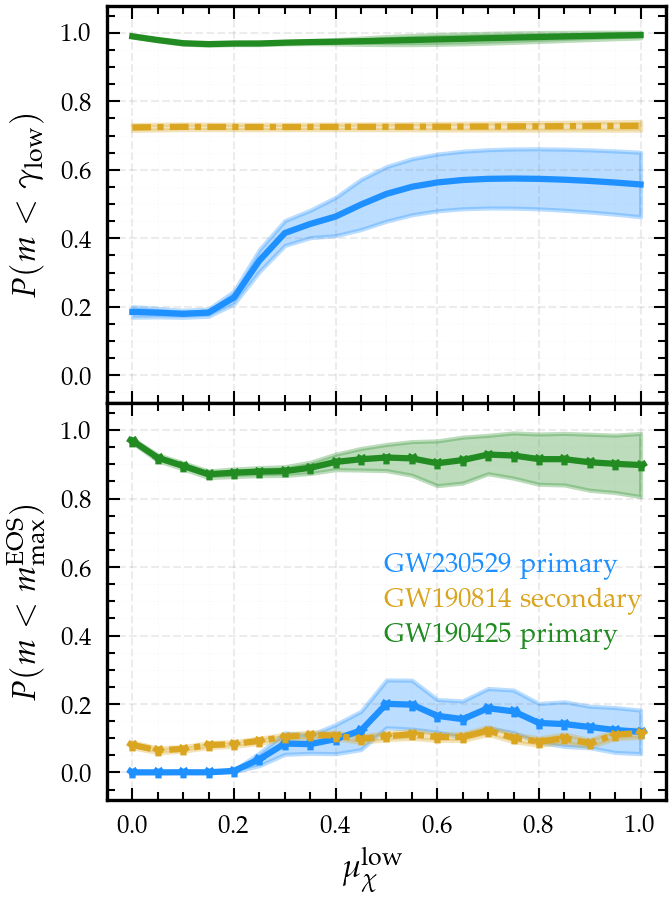}
\caption{\label{fig:muchi} Effect of the NS mean spin magnitude \muchi on classification.~\topbottompopeosmessage~\shededregionmessage}
\end{figure}

We next examine the coupling between mass ratio~\q~ and effective spin~\Xeff, focusing on how spin properties influence~\PNS. 
To disentangle these effects, we treat spin tilts and spin magnitudes separately. 
Figure~\ref{fig:mucostilt} varies the mean spin tilt~\mutilt, while Figures~\ref{fig:muchi} and \ref{fig:sigchi} explore the mean spin magnitude~\muchi~and its width~\sigchi, respectively. 
As summarized in Table~\ref{tab:feature_summary}, spin tilts produce comparable shifts to~\PNS~as spin magnitudes with the most pronounced responses arising in low SNR events where the \qXeff~degeneracy is broad.

\subsection{Spin tilts strongly vary~\PNS\ and shift primaries and secondaries in opposite directions \label{sec:qXeff_tilt}}
The Gaussian component of the spin tilt mixture, characterized by its mean~\mutilt~and width~\sigtilt, may influence the apparent coupling between the mass ratio~\q~and effective spin~\Xeff, which is determined by the likelihood.
Figures~\ref{fig:correlationscartoon}~and~\ref{fig:mucostilt}~illustrate this effect: increasing~\mutilt~toward completely aligned ($+1$) favors larger~\Xeff. 
Because~\q\ and~\Xeff\ are anticorrelated, increasing~\Xeff~pushes the reweighted posterior toward more asymmetric mass ratios (smaller~\q),  raising~\mone\ and lowering~\mtwo\ (see Figure~\ref{fig:correlationscartoon}). 

The classification response in Figure~\ref{fig:mucostilt} follows directly from this.
For primaries, greater alignment (larger~\mutilt) drives~\mone\ upward, pushing it past the NS boundary and lowering~\PNS.
For secondaries, greater alignment (larger~\mutilt) drives~\mtwo\ downward, pulling it deeper into the NS range and raising~\PNS. 

\subsection{Spin magnitudes also strongly vary~\PNS \label{sec:qXeff_mag}} 
The effect of varying the mean spin magnitude~\muchi\ for NS containing binaries is shown in Figure~\ref{fig:muchi}. 
Compared to~\mutilt, the impact on~\PNS~is also strong. 
The clearest trend arises in GW230529's primary. 
Increasing~\muchi~raises~\Xeff, which couples to~\q~through the~\qXeff\ anti-correlation, leading to higher~\PNS. 

The EOS-informed panel of Figure~\ref{fig:muchi} confirms this picture, with a strong feature seen in GW190425's primary. 
Here, the combination of low SNR and masses lying close to~\mmaxeos\ makes~\PNS\ sensitive to~\muchi. 
At very low~\muchi, the effective spin is driven toward $0$, which pushes~\q~toward symmetry, raising~\PNS. 
However, once~\muchi~is allowed to deviate from $0$,~\Xeff~rises, which drives~\q\ to smaller values and correspondingly lowers~\PNS. 
Figure~\ref{fig:muchi}~also shows that, at high~\muchi, low SNR events exhibit broader classification uncertainties. This arises from limited posterior support in the high spin magnitude subspace, where the scarcity of samples inflates the credible intervals. 

\begin{figure}[h!]
\includegraphics[width=\gridscale\columnwidth]{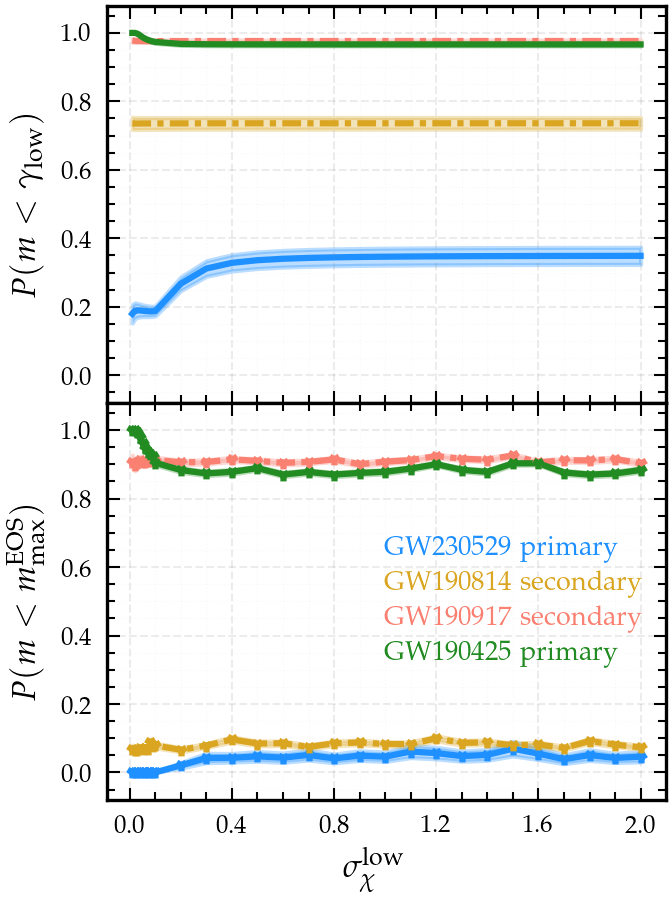}
\caption{\label{fig:sigchi} Effect of NS spin magnitude width, \sigchi on classification~\PNS.~\topbottompopeosmessage~\shededregionmessage}
\end{figure}

We now vary the width of the spin magnitude distribution~\sigchi. Across both the population-only and EOS-informed analyses, the resulting classifications are generally flat, with the only notable deviation occurring for GW190425 in the EOS panel when spin magnitudes are sharply peaked. 
For all other sources, broadening or narrowing~\sigchi~leaves~\PNS~essentially unchanged (see Figure~\ref{fig:sigchi}).

Taken together, Figures~\ref{fig:mucostilt}–\ref{fig:sigchi} show that spin magnitudes and tilts have a comparable influence on classification outcomes. Both sets of parameters can shift \PNS\, especially when event posteriors remain broad. 
Adjusting~\mutilt~can shift~\PNS~substantially (ranging from $\GWtwothreezerofivetwonineprimarymucostiltminPNS$\% to $\GWtwothreezerofivetwonineprimarymucostiltmaxPNS$\% for GW230529's primary and from $\GWoneninezerofourtwofiveprimarymucostiltEOSminPNS$\% to $\GWoneninezerofourtwofiveprimarymucostiltEOSmaxPNS$\% for GW190425's primary), with the largest changes occurring in low SNR sources.  
Spin magnitude (\muchi~and~\sigchi) produce similar shifts, most clearly visible in the population-only GW230529's primary (ranging from $\GWtwothreezerofivetwonineprimarymuchioneminPNS$\% to $\GWtwothreezerofivetwonineprimarymuchionemaxPNS$\%) and the EOS-informed GW190425's primary (ranging from $\GWoneninezerofourtwofiveprimarymuchioneEOSminPNS$\% to $\GWoneninezerofourtwofiveprimarymuchioneEOSmaxPNS$\%). 
Owing to its weaker observational constraints, \mutilt\ represents a more plausible driver of changes in \PNS\ than \muchi~\cite{collaborationGWTC40PopulationProperties2025}. For this reason, we choose \mutilt\ as the (marginally) more important spin parameter in assessing classification sensitivity.

\subsection{Jointly changing \bm{$\mutilt \text{ and } \pairingBNS$} maximally varies~\bm{\PNS} \label{sec:gowild}} 
Having examined each parameter in isolation, we now ``throw caution to the wind'' and simultaneously change two hyperparameters that strongly impact classification: the NS spin tilt mean (\mutilt) and the BNS pairing exponent (\pairingBNS). 
Specifically, we compare two extrema for primaries: \pairingBNS$=5$, \mutilt$=-1$ versus \pairingBNS$=-5$, \mutilt$=1$. 
For secondaries, we flip the sign of~\pairingBNS, because the $(q,\Xeff)$–$m_2$ correlation has the opposite sign to the $(q,\Xeff)$–$m_1$ correlation.
These two choices approximately bound the largest possible change to~\PNS, and the impact is dramatic. GW190425's primary sweeps from~\resulteosgwonenineofourtwofive, representing the largest shift observed across all EOS-informed cases. 
Under the population-only case, the GW230529 primary exhibits the greatest variance, spanning~\resultpopgwtwothreezerofiveonine. Similarly, the EOS case for GW230529 still permits a substantial deviation of \GWtwothreezerofivetwonineonebetamutiltcombinedEOSmin\% to \GWtwothreezerofivetwonineonebetamutiltcombinedEOSmax\%. 
By construction, these settings align (or misalign) spins while simultaneously favoring equal-mass (versus asymmetric pairings). These extremes set an envelope on the impact of uncertainty in the population. 

\section{\label{sec:lookingahead} Looking Ahead}
\begin{figure}[h!]
\includegraphics[width=0.99\columnwidth]{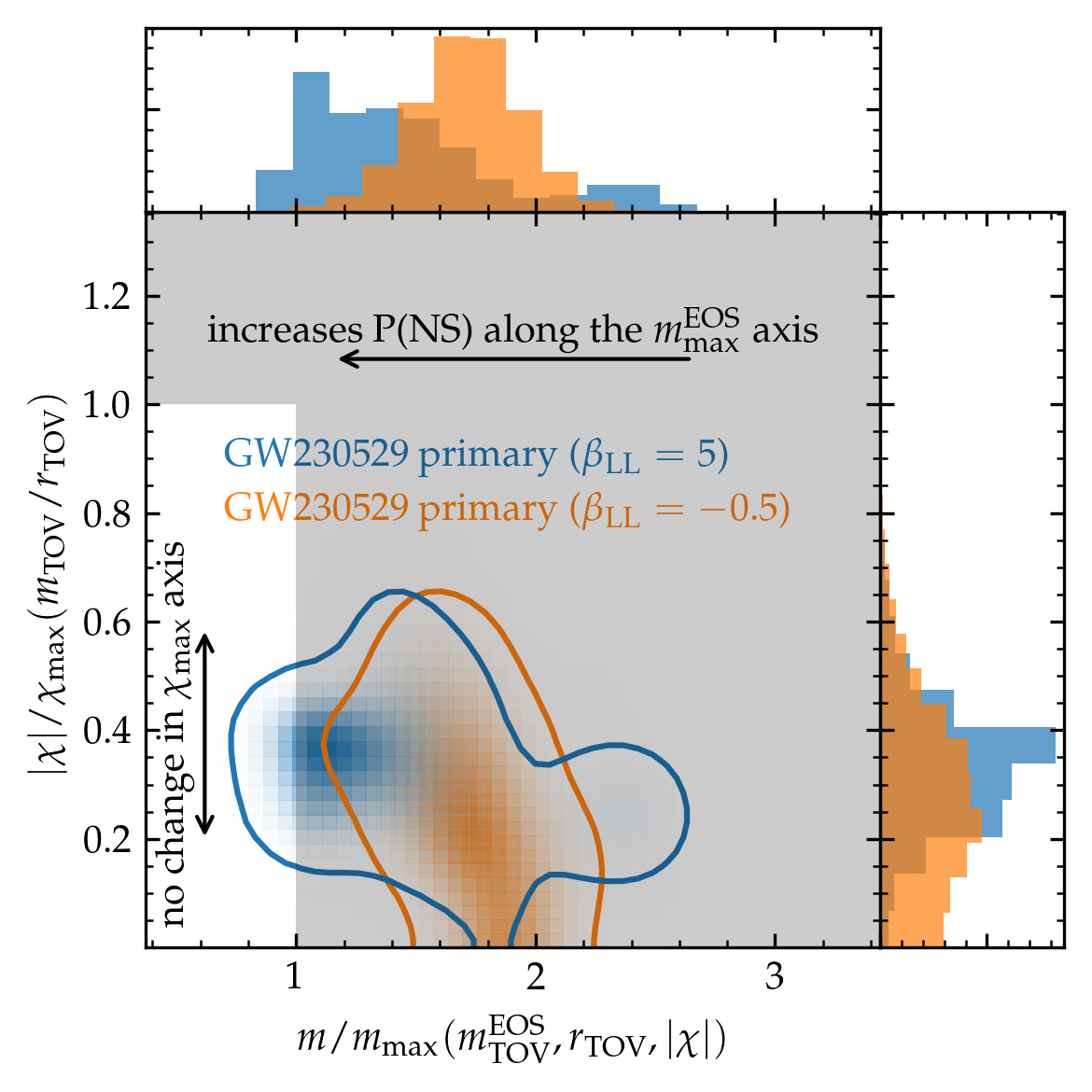}
\caption{\label{fig:mass-spin} EOS-informed reweighting for GW230529, comparing a strong preference for equal mass pairing ($\pairingBNS=5$, blue) with a weak preference for equal mass pairing ($\pairingBNS=-0.5$, orange). Both hyperparameter choices are equally likely. Reweighting shifts support along the $m/m_{\rm max}$ axis, moving the support from shaded ``Not NS'' region into the ``NS'' region. In contrast, little change is seen along the $|\chi|/\chi_{\rm max}$ axis, indicating that variations in~\PNS~are largely influenced through the EOS-informed maximum-mass rather than spin. }
\end{figure}
For the EOS-dependent case, NS classification near the NS-BH boundary is inherently two-dimensional. 
The physical boundary is set jointly by the EOS-dependent maximum mass and its spin (which can affect the maximum mass through~Equation \ref{eq:mmax}). 
Hence, we operate in the joint (\mmaxeos, $\chi_{\rm max}^{\rm EOS}$) plane rather than just the one dimensional $\mmaxeos$ cut. 
This representation shows which dimension (mass or spin) has a stronger effect on~\PNS. Reweighting primarily shifts probability along the $m/\mmaxeos$ axis, moving support from the shaded ``Not NS'' region into the ``NS'' region, while producing little change along $\chi/\chi_{\max}$. 
For realistic EOS, \mmaxeos~increases only weakly with spin, and the observed NS spins are either small or recover the prior (see Figure~\ref{fig:popspin}).
As a result, we do not expect the spins to greatly alter the reweighted population, exactly what is observed in Figure~\ref{fig:mass-spin}. 
For the maximum spin to have an effect, we will need to measure NS mergers with much larger spins. 

Having varied the population while holding the distribution over NS EOS fixed (using draws from \citet{landryNonparametricConstraintsNeutron2020}), a natural next step is to test different families of candidate EOS as a distribution over its physics~\cite{ finchUnifiedNonparametricEquationofstate2025, landryNonparametricConstraintsNeutron2020, ngInferringNeutronStar2025, golombInterplayAstrophysicsNuclear2025a, mohantyAstrophysicalConstraintsNeutron2024, douchinUnifiedEquationState2001, friedmanHotColdNuclear1981, lalazissisNewParametrizationLagrangian1997, wiringaEquationStateDense1988}. 
This framework could extend the analysis of \citet{essickDiscriminatingNeutronStars2020}, which examined GW190814 by systematically varying the EOS to assess whether it's secondary could be consistent with a NS (see their Table~2). 
Such studies are motivated by current constraints leaving a wide plausible range of mass--radius relations, which directly influence the maximum NS mass and, consequently, the value of~\PNS.

Beyond varying the EOS itself, another valuable extension could be to examine how alternative approaches to identifying the low-mass boundary of the population influence classification. For a first attempt, see Appendix~\ref{app:nonparam}.

\utty{It is also important to distinguish variability in~\PNS~that is supported by the data from the population. 
Several hyperparameters governing the low-mass and spin distributions are only weakly constrained by the data. Consequently, it is appropriate to evaluate~\PNS~across the full prior range. 
However, for parameters that are more tightly constrained, spanning the entire prior range can artificially inflate the apparent variability in~\PNS. Table~\ref{app:PCItable}~illustrates the resulting impact.
While adopting more restrictive support for~\PNS~would reduce the width of ranges quoted in Table~\ref{tab:fullresults}, it would not alter the conclusions of our work. 
The qualitative sensitivity of~\PNS~to population assumptions therefore remains robust.}

With relatively few BNS detections, the population of low-mass mergers remains fairly uncertain and variability in~\PNS\ remains broad. 
As detections increase in forthcoming observing runs, high SNR (and asymmetric) mergers will better constrain the NS population distribution, which will reduce the variability of~\PNS~for all events.

\section{\label{sec:conclusion} Conclusion}
NS classification from GW observations is governed by assumptions about the CBC population, the EOS, and single-event measurement uncertainties. 
In both the population-only and EOS-informed analyses, the low mass compact object pairing (\pairingBNS),  the NS spin tilt distribution (\mutilt) and the spin magnitude distribution (\muchi~and \sigchi) are the dominant parameters in shifting the inferred NS probability,~\PNS. By comparison, other hyperparameters move~\PNS\ moderately. 
Low SNR sources near the NS-BH boundary are the most sensitive to changes in classification. 
For example, under the population-only analysis, GW230529's primary classification probability spans \resultpopgwtwothreezerofiveonine, demonstrating the joint effects of \pairingBNS\ and \mutilt.
Using an EOS-informed approach, GW190425's primary varies from \resulteosgwonenineofourtwofive.
However, high-SNR asymmetric systems like GW190814 yield a~\PNS~that changes by $\leq 10\%$, indicating robustness to assumptions in the CBC population.

Looking ahead, three directions flow naturally. 
First, place greater confidence in classifications derived from low-mass, high-SNR signals, where the tighter posteriors reduce sensitivity to prior assumptions.
Second, when reporting classification metrics such as~\PNS, include an explicit uncertainty budget that surveys many population priors. 
Third, quantify how alternative EOS models modify~\PNS\ by repeating the analysis under multiple EOS prescriptions, for many events.
Collectively, these steps promote more robust classifications near the low-mass gap boundary

Ultimately, reliable NS classifications depend on confronting population model dependencies head on. As multi-detector observations of NS mergers accumulate, they will naturally refine the demographics of spin magnitudes, tilt angles, and mass ratios, and map detection rates across the low-mass gap. With these advances,~\PNS\ will enable more confident interpretation of future GW observations. 

\begin{acknowledgments}
U.M. thanks Aditya Vijayjumar, Sylvia Biscoveanu and Claire Ye for useful discussions.

U.M. and R.E thank Mike Zevin for reviewing the manuscript within the LVK. 

U.M. and R.E are supported by the Natural Sciences \& Engineering Research Council of Canada (NSERC) through a Discovery Grant (RGPIN-2023-03346).

This material is based upon work supported by NSF’s LIGO
Laboratory which is a major facility fully funded by the
National Science Foundation.

We also wish to acknowledge the land on which the University of Toronto operates. For thousands of years it has been the traditional land of the Huron-Wendat, the Seneca, and the Mississaugas of the Credit. Today, this meeting place is still the home to many Indigenous people from across Turtle Island and we are grateful to have the opportunity to work on this land.
\end{acknowledgments}

\onecolumngrid
\appendix
\section{\label{app:popmodel}Population Model}
We model the population distribution as a product of contributions from the component masses, spins, and redshift. 
Our framework extends the \textsc{FullPop4.0} model \cite{collaborationGWTC40PopulationProperties2025}, which builds on earlier formulations introduced by \citet{farahBridgingGapCategorizing2022, fishbachDoesMatterMatter2020}, and \citet{ maliStrikingChordSpectral2024}. 
This extension incorporates flexible features in the mass spectrum (power laws, peaks, and notches), mass-dependent spin prescriptions, and a redshift distribution that fits the evolving star formation rate. 
Together, these ingredients provide a unified hierarchical model for the CBC population.

We begin by factorizing our model into contributions from the redshift distribution, the joint mass distribution, and the spin distributions: 
\begin{equation}
p(\theta|\Lambda) = p(z|\Lambda)p(m_1,m_2|\Lambda)p(s_2|m_2, \Lambda)p(s_1|m_1, \Lambda)
\end{equation}
We will now explain our mass, spin and distance models separately. 
\subsection{Joint Mass Distribution\label{app:jointmass}}
The joint mass distribution is expressed as a product of two mass distributions together with a pairing function,
\begin{figure*}[h]
\includegraphics[width=0.55\textwidth]{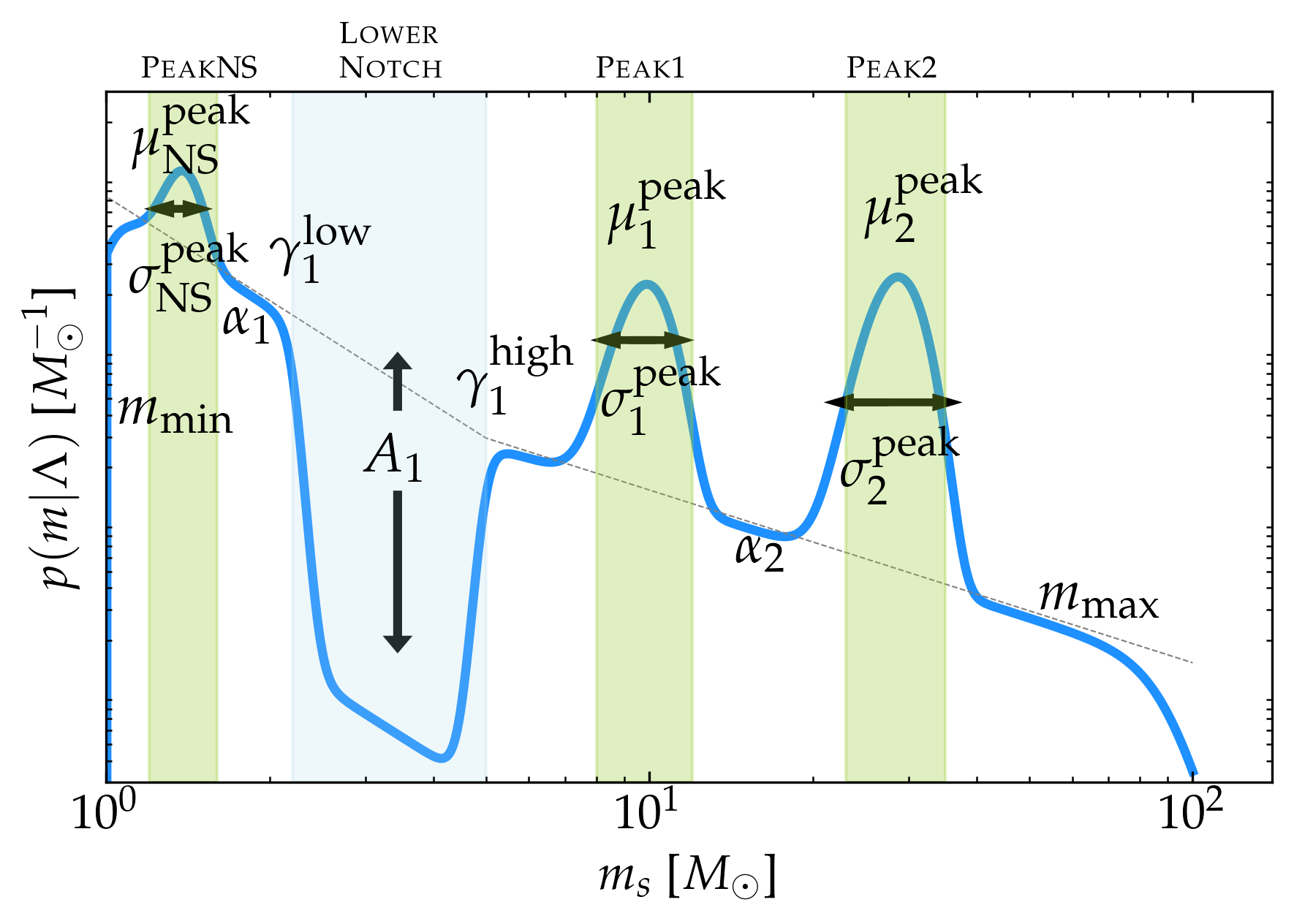}
\includegraphics[width=0.378\textwidth]{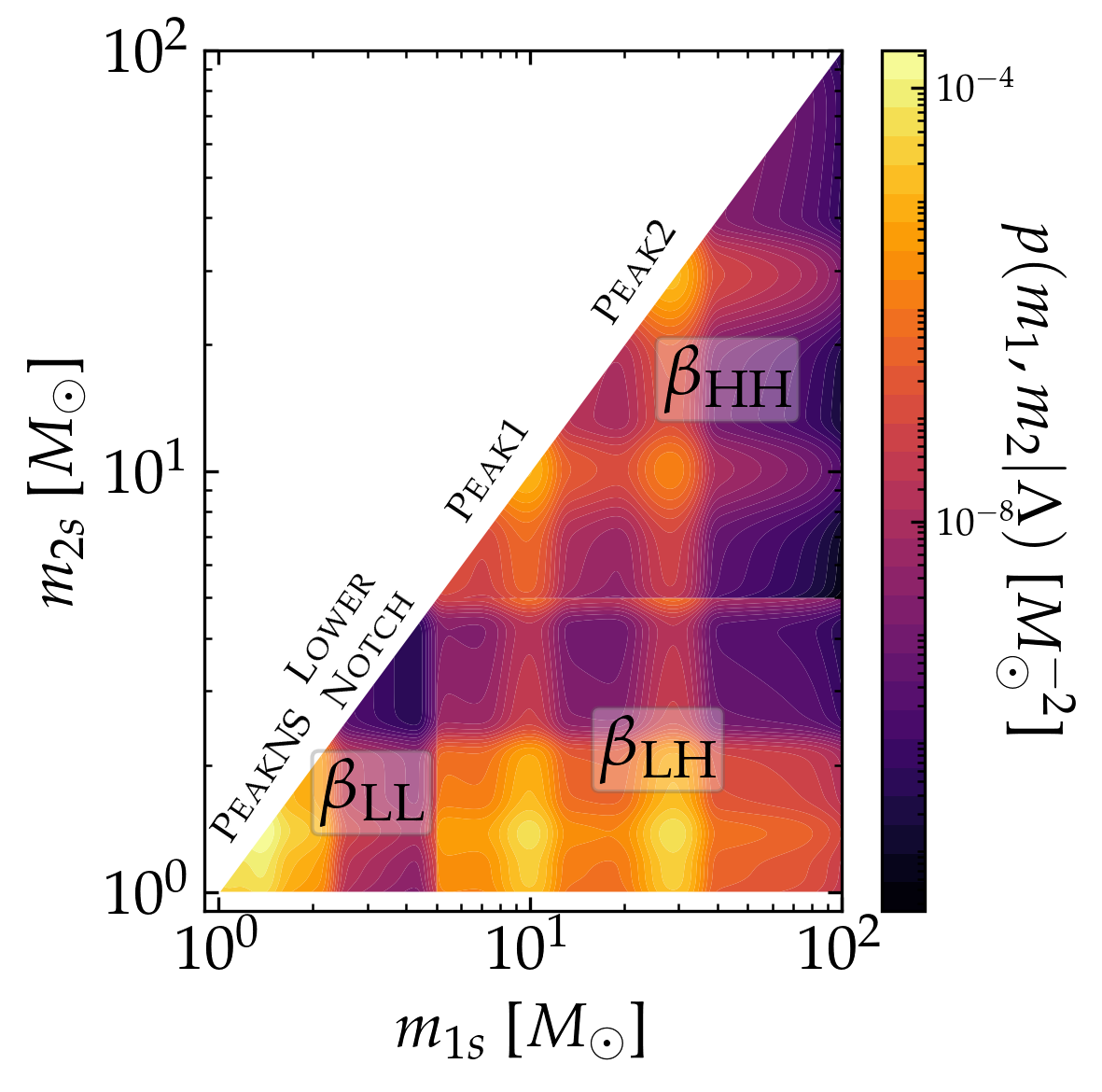}
\caption{\label{fig:pop_model} (\textbf{Left}) Illustration of the one-dimensional mass distribution $p(m|\Lambda)$ (Eq.~\ref{eq:1d-mass}) and its hyper-parameters. 
The baseline broken power law (\emph{dotted}) extends between~$m^{\rm model}_{\min}$ and $m^{\rm model}_{\max}$ (where the model tapers below $m_{\rm min}$ and above $m_{\rm max}$) with roll-offs $\gamma_{\rm low,high}$. 
Gaussian peaks (\emph{green}) represent localized excesses (e.g.\ NS peak, black hole peaks), while Butterworth notch filters (\emph{blue}) carve deficits. 
(\textbf{Right}) The corresponding joint distribution $p(m_{1},m_{2}|\Lambda)$ (Eq.~\ref{eq:joint mass}), where pairing hyper-parameters $\pairingBNS$, $\pairingNSBH$, and $\pairingBBH$ outline support across the Low-Low, Low-High, and High-High compact object ranges. 
Features in $p(m|\Lambda)$ imprint symmetrically on both $m_{1}$ and $m_{2}$. }
\end{figure*}
\begin{equation}
\label{eq:joint mass}
p(m_1,m_2|\Lambda) = p(m_1|\Lambda)p(m_2|\Lambda)f(m_{1}, m_{2}, \pairingBNS, \pairingBBH, \pairingNSBH)
\end{equation}
We consider a three-pronged pairing function with a separate pairing powerlaw for Low-Low, Low-High and High-High compact mergers (\pairingBNS, \pairingBBH, \pairingNSBH). 
The three pairing function is given by:

\begin{equation}\label{eq:pairing3}
f(m_{1s}, m_{2s}, \pairingBNS, \pairingBBH, \pairingNSBH) = 
    \begin{cases}
        \left(\frac{m_{2s}}{m_{1s}}\right)^{\pairingBNS}& \text{if}\ m_{2s} < 5 M_{\odot} \text{ and  } m_{1s} < 5 M_{\odot} \text{  (Low-Low)} \\
        \left(\frac{m_{2s}}{m_{1s}}\right)^{\pairingNSBH}& \text{if}\ m_{1s} \geq 5 M_{\odot} \text{ and  } m_{2s} \leq 5 M_{\odot} \text{  (Low-High)} \\
        \left(\frac{m_{2s}}{m_{1s}}\right)^{\pairingBBH}& \text{if}\ m_{2s} \geq 5 M_{\odot} \text{ and  } m_{1s} \geq 5 M_{\odot} \text{  (High-High)} \\
    \end{cases}
\end{equation}
The latent mass spectrum (in the source frame, i.e. $m_{1s}$,$m_{2s}$) incorporates a broken power law with flexible low and high mass cutoffs, Gaussian peaks, and notches (dips) to model over- and under-densities:
\begin{equation}
\begin{split}
p(m|\Lambda)
= \Big[&1
+ c_1 \,\mathcal{N}_{[m_{\min},\,m_{\max}]}\!\big(m \mid \mu_1,\sigma_1\big)
+ c_2 \,\mathcal{N}_{[m_{\min},\,m_{\max}]}\!\big(m \mid \mu_2,\sigma_2\big)
+ c_{\rm NS} \,\mathcal{N}_{[m_{\min},\,m_{\max}]}\!\big(m \mid \mu_{\rm NS},\sigma_{\rm NS}\big)
\Big] \\
&\quad \times~n \big(m|\gammalow,\gammahigh,\etalow,\etahigh,A_1\big)\ \\
&\quad \times~h \big(m|m_{\max},\eta_{\min}\big) \times \ell \big(m|m_{\max},\eta_{\max}\big)\ \\
&\quad \times~
\begin{cases}
\displaystyle \left(\dfrac{m}{\mbrk}\right)^{\alpha_1}\!, & m < \mbrk, \\[1.25ex]
\displaystyle \left(\dfrac{m}{\mbrk}\right)^{\alpha_2}\!, & m \ge \mbrk~,
\end{cases} \\
&\quad \times \begin{cases}
1, & m_{\min}^{\rm model} \leq m \leq m_{\max}^{\rm model} \\
0, & \text{otherwise}
\label{eq:1d-mass}
\end{cases}
\end{split}
\end{equation}
where $n(m)$ denotes the notch component, $h(m)$ and $\ell(m)$ are highpass and lowpass filters, and $\mathcal{N}_{[a,b]}$ are truncated Gaussians. 
Specifically, we use 

\begin{align}
    \ell(m|k, \eta) & = (1 + (m / k)^\eta )^{-1} \\
    h(m|k, \eta) & = 1 - \ell(m|k, \eta)
\end{align}
which smoothly turn the distribution off at characteristic scales $k$ with sharpness controlled by $\eta$.  
Localized dips or chips in the spectrum, such as the lower and upper notches shown in Figure~\ref{fig:pop_model}, are modeled through  
\begin{equation}
    n(m|\gammalow, \gammahigh, \etalow, \etahigh, A) = 1 - A \, h(m|\gammalow, \etalow) \, \ell(m|\gammahigh, \etahigh)
\end{equation}
where the parameters set the edges, widths, and depth of the dip. 

In addition to power law components and notches, we allow for Gaussian peaks. 
These are implemented with truncated normal distributions to ensure support only within specific bounds $[a,b]$:  
\begin{equation}
\mathcal{N}_{[a, b]}f(m \mid \mu, \sigma) =
\begin{cases}
\dfrac{\exp\!\left[-\tfrac{1}{2}\left(\tfrac{m-\mu}{\sigma}\right)^{2}\right]}
{\tfrac{1}{2}\,\sigma \sqrt{2\pi}\,\Bigg[\mathrm{erf}\!\Big(\tfrac{b-\mu}{\sqrt{2}\,\sigma}\Big) - \mathrm{erf}\!\Big(\tfrac{a-\mu}{\sqrt{2}\,\sigma}\Big)\Bigg]}, & a \leq m \leq b \\[2.0ex]
0, & \text{otherwise}
\end{cases}
\end{equation}
Tables~\ref{tab:pop_priors} and \ref{tab:pairing_priors} summarize the parameters governing the mass model and the pairing function. 
Figure~\ref{fig:pop_model} illustrates the structure of the mass model. 
The left panel shows how the low and high mass cutoffs, broken power law, Gaussian peaks, and notches combine. The right panel shows the resulting joint mass distribution, highlighting the role of the separate pairing functions for BNS, NSBH, and BBH sectors. 

\begin{table*}[t]
\centering
\caption{Summary of \popmodel\ mass population (and extensions) parameters and priors.}
\label{tab:pop_priors}
\begin{tabular}{@{} l l c p{9.5cm} c @{}}
\toprule
\textbf{Category} & \textbf{Parameter} & \textbf{Unit} & \textbf{Description} & \textbf{Prior} \\
\midrule
\multirow{3}{*}{Broken Power-Law}
 & $\alpha_1$  & --   & Power-law slope below \mbrk & $\mathrm{U}(-5,5)$ \\
 & $\alpha_2$  & --   & Power-law slope above \mbrk & $\mathrm{U}(-5,5)$ \\
 & \mbrk       & \Msun & Break point between $\alpha_1$ and $\alpha_2$ & $\mathrm{U}(2,10)$ \\
\midrule
\multirow{2}{*}{Highpass Filter}
 & $m_{\min}$  & \Msun & Low-mass roll-on location & $\mathrm{U}(1,1.2)$ \\
 & $\eta_{\min}$ & --  & Sharpness at $m_{\min}$ & $\mathrm{U}(10,50)$ \\
\midrule
\multirow{2}{*}{Lowpass Filter}
 & $m_{\max}$  & \Msun & High-mass roll-off location & $\mathrm{U}(35,100)$ \\
 & $\eta_{\max}$ & --  & Sharpness at $m_{\max}$ & $\mathrm{U}(0,10)$ \\
\midrule
\multirow{6}{*}{Notch (gap)}
 & \gammalow   & \Msun & Lower notch edge & $\mathrm{U}(2,4)$ \\
 & \etalow     & --    & Sharpness at \gammalow & $\mathrm{U}(0,50)$ \\
 & \gammahigh  & \Msun & Upper notch edge & $\mathrm{U}(4,8)$ \\
 & \etahigh    & --    & Sharpness at \gammahigh & $\mathrm{U}(0,50)$ \\
 & $A_1$         & --    & Lower Notch depth (dip amplitude; negative permits a bump) & $\mathrm{U}(-1,1)$ \\ 
\midrule
\multirow{3}{*}{Low-Mass Peak}
 & $\mu^{\rm peak}_1$     & \Msun & Peak location & $\mathrm{U}(6,12)$ \\
 & $\sigma^{\rm peak}_1$  & \Msun & Peak width & $\mathrm{U}(1,40)$ \\
 & $c_1$                  & --    & Peak height (amplitude) & $\mathrm{U}(0,500)$ \\
\midrule
\multirow{3}{*}{High-Mass Peak}
 & $\mu^{\rm peak}_2$     & \Msun & Peak location & $\mathrm{U}(20,60)$ \\
 & $\sigma^{\rm peak}_2$  & \Msun & Peak width & $\mathrm{U}(1,40)$ \\
 & $c_2$                  & --    & Peak height (amplitude) & $\mathrm{U}(0,500)$ \\
\midrule
\multirow{3}{*}{NS Peak}
 & \mupeakNS     & \Msun & Peak location & $\mathrm{U}(1.01,2.3)$ \\
 & \sigpeakNS    & \Msun & Peak width & $\mathrm{U}(0.1,1)$ \\
 & $c_{\rm NS}$  & --    & Peak height (amplitude) & $\mathrm{U}(0,1000)$ \\
\midrule
\multirow{2}{*}{Model Bounds}
 & $m_{\min}^{\rm model}$ & \Msun & Support lower bound & $1$ \\
 & $m_{\max}^{\rm model}$ & \Msun & Support upper bound & $100$ \\
\bottomrule
\end{tabular}
\end{table*}

\begin{table}[t]
\centering
\caption{Pairing function parameters and priors.}
\label{tab:pairing_priors}
\begin{tabular}{@{} l l c p{9.8cm} c @{}}
\toprule
\textbf{Category} & \textbf{Parameter} & \textbf{Unit} & \textbf{Description} & \textbf{Prior} \\
\midrule
\multirow{4}{*}{Pairing Function}
 & $\pairingBNS$        & --   & Spectral index for systems in the BNS range; ($m_1, m_2 < m_{\rm sep}$) & $\mathrm{U}(-5,5)$ \\
 & $\pairingBBH$        & --   & Spectral index for systems in the BBH range; ($m_1, m_2 > m_{\rm sep}$) & $\mathrm{U}(-5,5)$ \\
 & $\pairingNSBH$& --   & Spectral index for systems in the NSBH range; ($m_2 < m_{\rm sep}<m_1$) & $\mathrm{U}(-5,5)$ \\
 & $m_{\rm sep}$    & \Msun & Separator mass for pairing regimes & 5 \\
\bottomrule
\end{tabular}
\end{table}

\subsection{Mass-dependent Spin Distribution}
In addition to masses, the distribution of spin magnitudes and orientations plays a key role in the population. 
We model the spin degrees of freedom as conditionally independent given their component masses,
\begin{equation}
p(s_1,s_2|m_1,m_2,\Lambda)
=
\Big[p(\aone|m_1,\Lambda)\; p(\costiltone|m_1,\Lambda)\Big]\
\Big[p(\atwo|m_2,\Lambda)\; p(\costilttwo|m_2,\Lambda)\Big].
\end{equation}
This factorization separates the spin magnitudes ($\chi$) and cosine of the tilt angles ($\cos\theta$) and allows each to vary with the component mass. 
We implement a mass break at $\msb = 3\Msun$ to distinguish between objects in the low and high mass ranges. 
Below this threshold, spin magnitudes have a reduced maximum value due to the limited coverage of the O3 injections \cite{essickCompactBinaryCoalescence2025}.
The spin magnitude is split into a piecewise Gaussian,
\begin{equation}\label{eq:spinmag}
p(\chi \mid m,\Lambda) =
\begin{cases}
\TN{[\amin,\,\amaxNS]}\!\left(\chi \,\middle|\, \muchiOne, \sigchiOne\right), & m < \msb,\\[1.0ex]
\TN{[\amin,\,\amax]}\!\left(\chi \,\middle|\, \muchiTwo, \sigchiTwo\right),   & m \ge \msb,
\end{cases}
\end{equation}
where the means and widths of the truncated normals are separately parametrized above and below $\msb$.  
Similarly, the tilt distributions is mass-dependent.
Each is modeled as a mixture between an isotropic distribution and a truncated normal centered on alignment ($\cos\theta \approx 1$):
\begin{equation}\label{eq:spintilt}
p(\cos{\theta} \mid m,\Lambda) =
\begin{cases}
\mixtiltOne\;\TN{[\ctmin,\ctmax]}\!\left(\cos{\theta}|1, \sigtiltOne\right)
+ \left(1-\mixtiltOne\right)\;\frac{1}{\ctmax - \ctmin},
& m < \msb\\[1.0ex]
\mixtiltTwo\;\TN{[\ctmin,\ctmax]}\!\left(\cos{\theta}|1, \sigtiltTwo\right)
+ \left(1-\mixtiltTwo\right)\;\frac{1}{\ctmax - \ctmin},
& m \ge \msb
\end{cases}
\end{equation}
with $\ctmax = \cos{\theta_{\text{max}}} = +1$ and $\ctmin = \cos{\theta_{\text{min}}} = -1$ defining the support.  

Table~\ref{tab:spin_priors} summarizes the full set of hyperparameters and priors for the spin model. 

\begin{table*}[t]
\centering
\caption{Summary of mass dependent spin distribution parameters and priors. We denote a uniform distribution between X and Y as U (X, Y ).}
\label{tab:spin_priors}
\begin{tabular}{@{} r c p{8.0cm} c @{}}
\toprule
\textbf{Category} & \textbf{Parameter} & \textbf{Description} & \textbf{Prior} \\
\midrule
\multirow{4}{*}{Spin magnitude (\(m<\msb\))}
 & \muchiOne   & Mean of truncated normal below \(\msb\) & \(\mathrm{U}(0,\,0.4)\) \\[0.2em]
 & \sigchiOne  & Std.\ dev.\ of truncated normal below \(\msb\) & \(\mathrm{U}(0.05,\,2)\) \\[0.2em]
  & \amaxNS     & Upper bound on \(\chi_i\) & \(0.4\) \\
   & \amin       & Lower bound on \(\chi_i\) & \(0\) \\
\midrule
\multirow{4}{*}{Spin magnitude (\(m\ge\msb\))}
 & \muchiTwo   & Mean of truncated normal above \(\msb\) & \(\mathrm{U}(0,\,1)\) \\[0.2em]
 & \sigchiTwo  & Width of truncated normal above \(\msb\) & \(\mathrm{U}(0.05,\,2)\) \\[0.2em]
 & \amax       & Upper bound on \(\chi_i\) & \(1\) \\
  & \amin       & Lower bound on \(\chi_i\) & \(0\) \\
\midrule
\multirow{2}{*}{Tilt (\(m<\msb\))}
 & \mixtiltOne & Mixture fraction for truncated normal component& \(\mathrm{U}(0,\,1)\) \\[0.1em]
 & \sigtiltOne & Width of truncated normal component& \(\mathrm{U}(0.1,\,4)\) \\[0.1em]
\midrule
\multirow{2}{*}{Tilt (\(m\ge\msb\))}
 & \mixtiltTwo & Mixture fraction for truncated normal component& \(\mathrm{U}(0,\,1)\) \\[0.1em]
 & \sigtiltTwo & Width of truncated normal component& \(\mathrm{U}(0.1,\,4)\) \\[0.1em]
\midrule
\multirow{2}{*}{Spin tilt support}
 & \ctmax      & Upper bound on \(\cos\theta\) & \(1\) \\
 & \ctmin      & Lower bound on \(\cos\theta\) & \(-1\) \\
\midrule
\multirow{1}{*}{Spin mass break}
 & \msb        & Mass threshold separating the two spin regimes & \(3\Msun \) \\
\bottomrule
\end{tabular}
\end{table*}

\subsection{Redshift Distribution}
The final component of the population model describes the distribution of sources in redshift. 
This term accounts for the cosmological volume element, cosmological redshift, and rate-evolution as a function of cosmic time. We parameterize it as  
\begin{equation}\label{eq:p of z}
    p(\z \mid H_0,\Omega_{m,0},w,\kappaZ) 
    \propto \frac{dV_c}{d\z}(H_0,\Omega_{m,0},w)\,
    \frac{1}{1+\z}\,(1+\z)^{\kappaZ}
\end{equation}
where the comoving element is defined as
\begin{equation}\label{eq:dVcdz}
    \frac{dV_c}{d\z}(\z \mid H_0,\Omega_{m,0},w) 
    = \frac{4\pi\,c}{H_0}\,
      \frac{d_L^2(\z)}{E(\z)\,(1+\z)^2}
\end{equation}
and the factor of $(1+z)^{-1}$ in Equation~\ref{eq:p of z}~accounts for cosmological redshift, and the additional $(1+z)^{\kappaZ}$ term allows for evolution in the merger rate. 
This form makes explicit the dependence on cosmology through $H_0$, the matter density $\Omega_{m,0}$, and the dark-energy equation of state $w$, defined as the ratio of pressure to energy density for dark energy. The function \(E(\z)\) encapsulates how the cosmic expansion rate evolves with redshift and is determined by the fractional density parameters of the Universe today: matter (\(\Omega_{m,0}\)), radiation (\(\Omega_{r,0}\)), spatial curvature (\(\Omega_{k,0}\)), and dark energy (\(\Omega_{\Lambda,0}\)). 

Together, these parameters determine the comoving volume. Table~\ref{tab:distance_priors} lists the parameters used in this redshift model. 
In our analysis, all cosmological parameters are fixed to the Planck 2015 values \cite{collaborationPlanck2015Results2016}, while the redshift-evolution index $\kappaZ$ is left free. 
\begin{table}[t]
\centering
\caption{Summary of redshift distribution parameters and cosmology. All cosmology parameters are fixed to Planck 2015.}
\label{tab:distance_priors}
\begin{tabular}{@{} l l p{4.5cm} c @{}}
\toprule
\textbf{Category} & \textbf{Parameter} & \textbf{Description} & \textbf{Prior} \\
\midrule
Rate Evolution
 & \kappaZ & Redshift-evolution index & $\mathrm{U}(-4,8)$ \\
\midrule
\multirow{6}{*}{Cosmology}
 & $H_0$         & Present day expansion rate & 67.66 [$\text{kms}^{-1}\text{Mpc}^{-1}$]\\
 & $\Omega_{m,0}$    & Matter density & 0.311 \\
 & $\Omega_{r,0}$    & Radiation density & 0.001 \\
 & $\Omega_{k,0}$    & Curvature density & 0 \\
 & $\Omega_{\Lambda,0}$ & Cosmological constant density & $1 - \Omega_{m,0} - \Omega_{r,0} - \Omega_{k,0}$ \\
 & $w$           & Dark-energy equation-of-state  & $-1$ \\
\bottomrule
\end{tabular}
\end{table}

\section{GWTC-3.0 vs. GWTC-4.0 \label{app:GWTC3vs4}}
\utty{While GWTC--4.0 introduces new BHNS events that could affect the pairing and spin inference, it does not add confirmed BNS events. 
The dominant hyperparameters for~\PNS\ in our framework are the BNS pairing at low masses (\pairingBNS) and the NS tilt mean (\mutilt).
Consequently, the parts of the population that control the NS mass, spin, and redshift distributions remain constrained by the same information already present in GWTC--3.0.  
Although GWTC--4.0 highlights modest updates to the BH spin distributions \cite{collaborationGWTC40PopulationProperties2025}, our population treats BH and NS with separate spin magnitude and tilt hyperparameters. 
As a result, we do not expect a great change to \mutilt, nor do we expect new features at low masses that would qualitatively alter our conclusions. 
For these reasons, we expect that updating our population inference to include GWTC--4.0 would not alter our main conclusions. 
We emphasize that the goal of this work is not to provide a definitive estimate of~\PNS~for any individual event, but to demonstrate that~\PNS~can vary substantially under reasonable population assumptions. 
}

\section{\label{app:gw230529} Comparisons to the GW230529 exceptional event paper}

When conducting our analysis, we attempted to reproduce the published GW230529 results from~\citet{collaborationObservationGravitationalWaves2024} and encountered some discrepancies. 
A thorough review revealed an error in the original classification code, which prevented the population parameters from being correctly updated when marginalizing over the hyperposterior.
This only affects the quoted probabilities for the \textsc{Power law + Dip + Break} model in Table 3 and the surrounding discussion in~\citet{collaborationObservationGravitationalWaves2024}.
Correcting for this, we find that $P(m_1\text{ is NS})$ drops from $8.8 \pm 2.8\%$ to $4.2 \pm 0.7\%$ and $P(m_2\text{ is NS})$ increases from $98.4 \pm 1.3\%$ to $99.2 \pm 0.4\%$ when using the default spin assumptions (isotropic orientations and uniform in magnitude up to a maximum of $0.4$).
When these spin assumptions are relaxed and spin magnitudes can be as large as $0.99$, we now find that $P(m_1\text{ is NS})$ only increases to $13.1 \pm 0.7\%$ instead of to $27.3 \pm 3.8\%$.

While the exact classification probabilities change, they do not affect the key takeaways from~\citet{collaborationObservationGravitationalWaves2024}.
Additionally, our analysis explores a much broader range of possible populations, and we find that the primary can still reach probabilities as large as $\mathcal{O}$(1 in 2) in population-only analysis (see Section~\ref{sec:gowild}) and $\mathcal{O}$(1 in 4) under specific EOS-informed choices (see Section~\ref{sec:qm1m2_pairing}~and~\ref{sec:qXeff_tilt}). 
While such conclusions should not be overinterpreted, they nonetheless reinforce the original point presented in~\citet{collaborationObservationGravitationalWaves2024} that event classifications can depend strongly on poorly constrained aspects of the population of merging binaries.

\section{\label{app:mmmsextra} Classification of Neutron Stars - Monte Carlo Uncertainty}

\begin{table}[t]
\centering
\renewcommand{\arraystretch}{1.25}
\setlength{\tabcolsep}{10pt}
\caption{\label{app:PCItable}\utty{Summary of the inferred population hyperparameters used to compute~\PNS. We report 90\% posterior credible intervals from the  population inference, together with the full prior ranges over which each parameter is varied when evaluating \PNS. These prior ranges define the parameter sweeps used in Tables~\ref{tab:feature_summary} and~\ref{tab:fullresults}.}}
\begin{tabular}{l l cc}
\toprule
\textbf{Parameter} & \textbf{Description} & \textbf{90\% Credible Interval} & \textbf{Prior range of P(NS)} \\
\midrule
$\pairingBNS$ &
Pairing between BNS &
$\betaonemin$ -- $\betaonemax$ &
$\betaonePNSmin$ -- $\betaonePNSmax$ \\
$\mupeakNS$ &
Peak in the NS mass distribution &
$\mupeakNSmin$ -- $\mupeakNSmax$ &
$\mupeakNSPNSmin$ -- $\mupeakNSPNSmax$ \\
$\sigpeakNS$ &
Width of the NS mass peak &
$\sigpeakNSmin$ -- $\sigpeakNSmax$ &
$\sigpeakNSPNSmin$ -- $\sigpeakNSPNSmax$ \\
$\mutilt$ &
Spin tilt distribution of NS &
$\mucostiltmin$ -- $\mucostiltmax$ &
$\mucostiltPNSmin$ -- $\mucostiltPNSmax$ \\
$\sigtilt$ &
Width of the NS spin tilt distribution &
$\sigtiltonemin$ -- $\sigtiltonemax$ &
$\sigtiltonePNSmin$ -- $\sigtiltonePNSmax$ \\
$\muchi$ &
Spin magnitude distribution of NS &
$\muchionemin$ -- $\muchionemax$ &
$\muchionePNSmin$ -- $\muchionePNSmax$ \\
$\sigchi$ &
Width of the NS spin magnitude &
$\sigchionemin$ -- $\sigchionemax$ &
$\sigchionePNSmin$ -- $\sigchionePNSmax$ \\
$\gammahigh$ &
Upper edge of the NS--BH dip &
$\gammahighmin$ -- $\gammahighmax$ &
$\gammahighPNSmin$ -- $\gammahighPNSmax$ \\
\bottomrule
\end{tabular}
\end{table}

We now describe the Monte Carlo sums used to estimate~\PNS~throughout our study. 
For individual events with data $d_i$, we use posterior samples constructed using a reference population model $\Lambda_{\text{ref}}$:
\begin{equation}
p(\theta | d_i, \Lambda_{\text{ref}}) = \frac{p(d_i | \theta) \, p(\theta | \Lambda_{\text{ref}})}{p(d_i | \Lambda_{\text{ref}})}
\label{eq:event_posterior}
\end{equation}

We then use importance sampling to evaluate integrals over $\theta$. Given $M$ samples $\theta_k$ drawn from $p(\theta|d, \Lambda_{\rm ref}, H)$ with weights $W_k$, the expectation value over a function $F(\theta)$ is approximated as:
\begin{equation}
\int d\theta \, p(\theta | d, \Lambda_{\text{ref}}, H) \, F(\theta) \approx \frac{1}{M} \sum_{k=1}^M W_k F(\theta_k)
\label{eq:mc_inner}
\end{equation}
When evaluating the same function under a different population model $\Lambda$, we reweight the samples via:
\begin{equation}
\int d\theta \, p(\theta | d, \Lambda, H) \, F(\theta) \approx
\left( \sum_{k=1}^M w_k(\Lambda) \right)^{-1}
\sum_{k=1}^M w_k(\Lambda) \, F(\theta_k)
\label{eq:mc_reweighted}
\end{equation}
with
\begin{equation}
\omega_k(\Lambda) = W_k \frac{p(\theta_k | \Lambda, H)}{p(\theta_k | \Lambda_{\text{ref}}, H)}
\label{eq:importance_weights}
\end{equation}
Referring back to Equation~\ref{eq:prob_ns}, the outer integral over $\Lambda$ and $\EOS$ is approximated with $N$ samples $(\Lambda_p, \epsilon_p)$ with weights $\omega_p$:
\begin{equation}
\begin{split}
& \int d\Lambda \, d\EOS \, p(\Lambda, \EOS | H) \, F(\Lambda, \EOS) \approx
\left( \sum_{p=1}^N \omega_p \right)^{-1} \sum_{p=1}^N \omega_p F(\Lambda_p, \EOS_p)
\label{eq:mc_outer}
\end{split}
\end{equation}
Combining these expressions, we estimate the probability that an object is consistent with a NS as:
\begin{equation}
P(\rm{NS}) =\frac{1}{\sum_p^N \omega_p} \sum_p^N\left[\frac{\omega_p}{\sum_k^M \omega_k\left(\Lambda_p\right)} \sum_k^M\left[\omega_k\left(\Lambda_p\right) \Theta_{\mathrm{NS}}\left(\theta_k \mid \varepsilon_p\right)\right]\right]
\label{eq:p_hat}
\end{equation}
where $\varepsilon_p$ is the inferred population ($\Lambda_p$) in the population-only case and the combined population in the EOS-informed case ($\Lambda_p, \epsilon_p$). 
We separate Eq.~\ref{eq:p_hat} into four pieces and compute the moments of each in turn.
Specifically, we define
\begin{equation}
    \hat{\mathcal{P}} = \frac{F}{G}
\end{equation}
where
\begin{align}
    G & = \frac{1}{N} \sum\limits_p^N \omega_p \\
    F & = \frac{1}{N} \sum\limits_p^N \omega_p \frac{f(\Lambda_p, \varepsilon_p)}{g(\Lambda_p)}
\end{align}
and
\begin{align}
    g(\Lambda) & = \frac{1}{M} \sum\limits_k^M w_k(\Lambda) \\
    f(\Lambda, \varepsilon) & = \frac{1}{M} \sum\limits_k^M w_k(\Lambda) \Theta_\mathrm{NS}(\theta_k|\varepsilon)
\end{align}
We compute moments (expectation values $\mathrm{E}[\cdot]_x$, variances $\mathrm{V}[\cdot]_x$, and/or covariances $\mathrm{C}[\cdot,\cdot]_x$) of each under the measures associated with drawing samples $x$ (either $\theta_k \sim p(\theta)$ and $\Lambda_p, \varepsilon_p \sim p(\Lambda,\varepsilon)$).
We then approximate the moments of the ratio assuming the uncertainty in each sum is small
\begin{align}
    \mathrm{E}\left[\frac{a}{b}\right]_x & \approx \frac{\mathrm{E}[a]_x}{\mathrm{E}[b]_x} \label{eq:mean ratio} \\
    \mathrm{V}\left[\frac{a}{b}\right]_x & \approx \frac{1}{\mathrm{E}[b]_x^2} \mathrm{V}[a]_x + \frac{\mathrm{E}[a]_x^2}{\mathrm{E}[b]_x^4} \mathrm{V}[b]_x - \frac{2\mathrm{E}[a]_x}{\mathrm{E}[b]_x^3} \mathrm{C}[a,b]_x \label{eq:var ratio}
\end{align}
which comes from approximating errors as Gaussian and expanding the ratio to first order in terms of small errors away from the expected values.


\subsection{Monte Carlo uncertainty from sums over single-event uncertainty}
\label{sec:single-event uncertainty}

Now, one can compute moments of Monte Carlo sums by integrating over the measure that defines how the samples were drawn
\begin{align}
    \mathrm{E}[g]_\theta & = \int \prod\limits_\kappa^M d\theta_\kappa \, p(\theta_\kappa) \left(\frac{1}{M} \sum_k^M w_k \right) \nonumber \\
        & = \int d\theta \, p(\theta) w(\theta, \Lambda)
\end{align}
Similarly, by computing the second moment of $g$, we obtain
\begin{equation}
    \mathrm{V}[g]_\theta = \frac{1}{M} \left( \int d\theta\, p(\theta) w(\theta,\Lambda)^2 - \left(\int d\theta\, p(\theta) w(\theta,\Lambda) \right)^2 \right)
\end{equation}
We can approximate both these moments with sums over the samples $\theta_k \sim p(\theta)$
\begin{align}
    \mathrm{E}[g]_\theta & \approx \frac{1}{M} \sum\limits_k^M w_k(\Lambda) \\
    \mathrm{V}[g]_\theta & \approx \frac{1}{M} \left( \left[\frac{1}{M}\sum_k^M w_k^2\right] - \mathrm{E}[g]_\theta^2 \right)
\end{align}
We also obtain estimates for the moments of $f$
\begin{align}
    \mathrm{E}[f]_\theta & \approx \frac{1}{M} \sum\limits_k^M w_k \Theta_k \\
    \mathrm{V}[f]_\theta & \approx \frac{1}{M} \left( \left[\frac{1}{M} \sum\limits_k^M w_k^2 \Theta_k\right] - \mathrm{E}[f]_\theta^2 \right)
\end{align}
where $\Theta_k = \Theta_\mathrm{NS}(\theta_k|\varepsilon)$ and
\begin{equation}
    \mathrm{C}[f,g]_\theta \approx \frac{1}{M} \left( \left[\frac{1}{M}\sum\limits_k^M w_k^2 \Theta_k\right] - \mathrm{E}[f]_\theta \mathrm{E}[g]_\theta \right)
\end{equation}

Noting that these moments depend on both $\Lambda$ and $\varepsilon$, one can approximate $\mathrm{E}[f/g]_\theta$ and $\mathrm{V}[f/g]_\theta$ for each sample $(\Lambda_p, \varepsilon_p)$ separately via Eqs.~\ref{eq:mean ratio} and~\ref{eq:var ratio}.


\subsection{Monte Carlo uncertainty from sums over population uncertainty}
\label{sec:pop uncertainty}

We now consider the uncertainty from the finite number of population samples.
Following a similar procedure as Sec.~\ref{sec:single-event uncertainty}, it is straightforward to show that
\begin{align}
    \mathrm{E}[G]_{\Lambda,\varepsilon} & \approx \frac{1}{N} \sum\limits_p^N \omega_p \\
    \mathrm{V}[G]_{\Lambda,\varepsilon} & \approx \frac{1}{N} \left( \left[\frac{1}{N}\sum\limits_p^N \omega_p^2\right] - \mathrm{E}[G]_{\Lambda,\varepsilon}^2 \right)
\end{align}
The moments of $F$ are slightly more complicated, but it is also possible to show that
\begin{equation}
    \mathrm{E}[F]_{\Lambda,\varepsilon,\theta} \approx \frac{1}{N} \sum\limits_p^N \omega_p \mathrm{E}\left[\frac{f_p}{g_p}\right]_\theta
\end{equation}
where $f_p = f(\Lambda_p, \varepsilon_p)$ and $g_p = g(\Lambda_p)$.
One can also show that
\begin{align}
    \mathrm{E}[F^2]_{\Lambda,\varepsilon,\theta}
        & \approx \frac{1}{N} \left( \frac{1}{N} \sum\limits_p^N \omega_p^2 \mathrm{E}\left[\left(\frac{f_p}{g_p}\right)^2\right]_\theta \right) + \left(1 - \frac{1}{N}\right) \frac{1}{N^2} \sum\limits_p^N \omega_p \sum\limits_q^N \omega_q \mathrm{E}\left[\frac{f_p}{g_p}\frac{f_q}{g_q}\right]_\theta
\end{align}
Expanding the moments with respect to $\theta$ in terms of expected values and (co)variances yields
\begin{align}
    \mathrm{V}[F]_{\Lambda,\varepsilon,\theta}
        & \approx \frac{1}{N} \left( \left[\frac{1}{N} \sum\limits_p^N \omega_p^2 \mathrm{E}\left[\frac{f_p}{g_p}\right]_\theta^2\right] - \mathrm{E}[F]_{\Lambda,\varepsilon,\theta}^2 \right) \nonumber \\
        & \quad + \left(2 - \frac{1}{N}\right) \frac{1}{N^2} \sum\limits_p^N \omega_p^2 \mathrm{V}\left[\frac{f_p}{g_p}\right]_\theta \nonumber \\
        & \quad + \left(1 - \frac{1}{N}\right) \frac{1}{N^2} \sum\limits_{p}^N \omega_p \sum\limits_{q\neq p}^N \omega_q \mathrm{C}\left[\frac{f_p}{g_p}\frac{f_q}{g_q}\right]_\theta \label{eq:var F}
\end{align}

Let us interpret each of these terms in turn.
The first term is the standard expression for the uncertainty in a Monte Carlo sum assuming the estimate $\mathrm{E}[f/g]_\theta$ is known exactly for each ($\Lambda, \varepsilon$) sample.
The second and third lines account for the additional uncertainty from the fact that $\mathrm{E}[f/g]_\theta$ is not a perfect estimator.
If one considers the second line by itself, this is proportional to the variance expected from adding $N$ uncorrelated noisy estimators together, each with a weight $\omega_p/N$.
This term will vanish as $O(1/N)$ as the number of population samples increases.
This is because, even if each individual estimator is noisy, the sum of a very large number of noisy estimators can still be a precise estimator.
The third line, however, does not (in general) vanish as $N\rightarrow\infty$.
This represents the residual uncertainty from the fact that we reuse the same single-event parameter samples for each population sample.
In fact, this term scales as $\mathrm{C}[f_p f_q / g_p g_q]_\theta \sim O(1/M)$.

One could estimate $\mathrm{C}[(f_p/g_p)(f_q/g_q)]_\theta$ directly for all pairs of population samples, but this scales as $N(N+1)/2$ and may be extremely costly for large sample sizes.
Instead, we consider the upper limit
\begin{equation}
    \left| \mathrm{C}\left[\left(\frac{f_p}{g_p}\right)\left(\frac{f_q}{g_q}\right)\right]_\theta \right| \leq \sqrt{\mathrm{V}\left[\frac{f_p}{g_p}\right]_\theta \mathrm{V}\left[\frac{f_q}{g_q}\right]_\theta}
\end{equation}
and obtain
\begin{align}
    \mathrm{V}&[F]_{\Lambda,\varepsilon,\theta} \lesssim \frac{1}{N} \left( \frac{1}{N} \sum\limits_p^N \omega_p^2 \mathrm{E}\left[\frac{f_p}{g_p}\right]_\theta^2 - \left(\frac{1}{N}\sum\limits_p^N \omega_p \mathrm{E}\left[\frac{f_p}{g_p}\right]_\theta\right)^2 \right) \nonumber \\
        & \quad + \frac{1}{N}\left( \frac{1}{N} \sum\limits_p^N \omega_p^2 \mathrm{V}\left[\frac{f_p}{g_p}\right]_\theta - \left(\frac{1}{N} \sum\limits_p^N \omega_p \mathrm{V}\left[\frac{f_p}{g_p}\right]_\theta^{1/2}\right)^2 \right) \nonumber \\
        & \quad\quad + \left( \frac{1}{N} \sum\limits_p^N \omega_p \mathrm{V}\left[\frac{f_p}{g_p}\right]_\theta^{1/2} \right)^2
\end{align}
Again, the first two lines vanish as $O(1/N)$, but the third line remains finite.
This persistent term is an averaged uncertainty from the individual estimates derived from sums over single-event parameter uncertainty, and it corresponds to the variance in the sum of many perfectly correlated variates.
Again, it will instead vanish as $O(1/M)$.

Finally, we can also approximate
\begin{equation}
    \mathrm{C}[F,G]_{\Lambda,\varepsilon,\theta} \approx \frac{1}{N} \left( \left[\frac{1}{N}\sum\limits_p^N \omega_p^2 \mathrm{E}\left[\frac{f_p}{g_p}\right]_\theta\right] - \mathrm{E}[G]_\Lambda \mathrm{E}[F]_{\Lambda,\varepsilon,\theta} \right)
\end{equation}

With these expressions, we can then approximate $\hat{\mathcal{P}}$ and its uncertainty via Eqs.~\ref{eq:mean ratio} and~\ref{eq:var ratio}.


The expressions above are implemented within the publicly available library \texttt{mmax-model-selection}~\cite{mmax-model-selection}, which in turn depends on \texttt{gw-distributions}~\cite{gw-distributions} for models of the astrophysical population.
These libraries are lightweight and, for $N \sim M \sim 1000$, estimates of the prior/posterior probabilities, odds ratios, and the Bayes factor can be run on a laptop (a single Apple M3 Pro CPU clocked at 4.05 GHz) in $O(15)$ sec.

\section{\label{app:nonparam} What about non-parametric mass models?}\label{app:nonparam:methods}

We intentionally do not deploy a non-parametric mass model here for two reasons. 
First, the core aim of this paper is interpretability: we wish to isolate how specific, physically meaningful population parameters (NS peak, pairing, spin magnitudes/tilts etc.) affect classification. Non-parametric models, while flexible, tend to absorb such effects into an overall feature making such connections opaque. 
Second, introducing a non-parametric layer to classification would add additional priors whose influence may rival the physics we seek to probe. 
This highlights a more general problem with the tradeoff between interpretable models which are highly parametrized and weakly parametrized models which lack interpretability.

\utty{From this perspective, many commonly adopted ``simpler'' population models can be understood as enforcing specific astrophysical assumptions within the same conceptual framework.
For example, assuming that the spin or mass distributions inferred from higher-mass black holes apply unchanged to lower-mass systems amounts to fixing population structure that is only weakly constrained by current data.
We have explored such simplified variants and find no change to our conclusions.
}

Nonetheless, if one wishes to calculate~\PNS~using a non-parametric mass model, we provide a proof-of-concept below. 
Such a method could estimate classification probabilities~\PNS~directly from $p(m|\Lambda)$, bypassing the need for explicit parameters representing the NS-BH boundary. 

We introduce a sample, two-stage edge detection procedure that scans the latent mass distribution for a sharp down turn consistent with the NS-BH boundary and report this as an estimate of the maximum non-rotating NS mass (which is then used to infer~\PNS). 
The algorithm operates on a mass grid $m \in [1,10]$ and requires a set of posterior draws from the population mass distribution $p(m|\Lambda)$ (or any posterior distribution on a similar grid). 

First, we apply a high frequency ``inflection'' filter, implemented as a second derivative (top hat) kernel, by convolving it with the mass function. 
This filter highlights abrupt curvature changes and edges. 
We selected the location of the peak filter response as our candidate \mtovdet. 
If the resulting candidate lies in a physically realistic range $2 \Msun < \mtovdet < 5 \Msun$, we accept it.
Otherwise, we repeat the search with a broader ``valley'' filter, a local minimum kernel tuned to capture wide, gently sloping transitions. 
After convolving the second filter with the mass function, we once again select its peak filter response. 

Since this method is purely filter based and works on $p(m|\Lambda)$, it is robust to specific parametric forms of the underlying population (i.e. this method works on any non-parametric mass model).

This is shown in the algorithm below. See \texttt{github.com/utkarsh7236/Guesswork-in-the-Gap} for the complete implementation. 
\begin{center}
\rule{\linewidth}{0.8pt}
\vspace{-0.8em}
\noindent\textbf{Algorithm 1}: Automatic detection of the NS--BH boundary\label{alg:cap}
\rule{\linewidth}{0.4pt}
\vspace{-1.5em}
\end{center}
\begin{minipage}{0.95\linewidth}
\begin{algorithmic}[1]
  \State \textbf{Input:} Posterior mass probability $p(m|\Lambda)$ on $m \in [1,10]\,M_\odot$
  \State Compute log density: $s(m) = \log_{10} p(m|\Lambda)$
  \State Convolve $s(m)$ with $K_1$ (inflection kernel) $\rightarrow r_1(m)$
  \State Find candidate $m^{\mathrm{candidate}} = \arg\max |r_1(m)|$
  \If{$2.0 < m^{\mathrm{candidate}} < 4.9$}
      \State Accept $\mtovdet = m^{\mathrm{candidate}}$
  \Else
      \State Convolve $s(m)$ with $K_2$ (valley kernel) $\rightarrow r_2(m)$
      \State Find $m^{\mathrm{candidate}} = \arg\max |r_2(m)|$
      \State Set $\mtovdet = m^{\mathrm{candidate}}$
  \EndIf
  \State Repeat steps 1--10 for all posterior draws to build the set $\{\mtovdet\}$
\end{algorithmic}
\vspace{-0.8em}
\rule{\linewidth}{0.4pt}
\end{minipage}
\label{app:nonparam:edge detection}

\begin{figure*}[t]
\includegraphics[width=\textwidth]{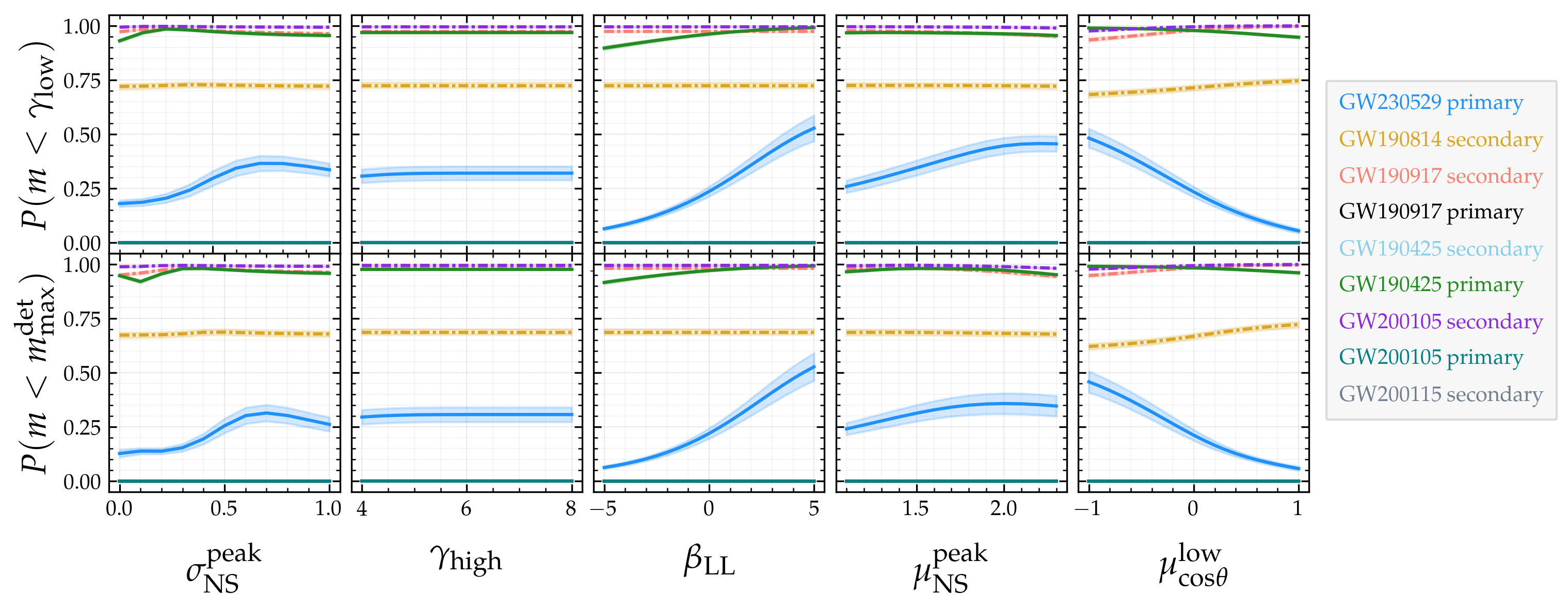}
\caption{\label{fig:nonparametric} (\textbf{Top}) Classification probabilities obtained with the parametric population model showing $P(m < \gammalow )$. (\textbf{Bottom}) the non-parametric method introduced in Appendix~\ref{app:nonparam} $P(m < \mmaxnonparam)$. Each panel varies one hyper-parameter. Across all parameters, the two approaches demonstrate remarkable agreement. Both identify the same similar changes in each event across the hyperparameter range. Overall, the consistency between methods validates our parametric results while demonstrating a proof-of-concept non-parametric approach (which can be applied to non-parametric mass models).~\shededregionmessage}
\end{figure*}

\vspace{0.8em}
Referring to Figure~\ref{fig:nonparametric}, we find that the trends inferred from the parametric and non-parametric edge detection are remarkably similar across all hyperparameters considered. 
In both cases, the inferred~\PNS~remain consistent with one another. 
This consistency demonstrates the feasibility of such an analysis on weakly modeled mass distributions.

This method for detecting features is not particularly well explored.
We have selected two detection schemes that, at best, may capture hints of structure in the mass distribution.
They depend on a few heuristic criteria that are far from reliable.
In principle, one could devise any number of similar algorithms to “discover” features marking the onset of the lower mass gap, each equally defensible.
We restrict ourselves to two, as they perform adequately for illustrative purposes.
Ultimately, these detection choices encode implicit priors and underscore the limitations of non-parametric approaches, which can often appear more objective than they actually are.

\begin{figure*}[t]
\includegraphics[width=\textwidth]{corner.pdf}
\caption{\label{fig:corner} \utty{Corner plot of the joint posterior distribution of population hyperparameters inferred from GWTC--3, illustrating correlations between parameters. Labeling follows the~\popmodel~defined as \textsc{MultiPDB}~in~\citet{maliStrikingChordSpectral2024}.}
}
\end{figure*}


\twocolumngrid
\clearpage
\bibliographystyle{unsrtnat}
\bibliography{mmms, zessick}
\end{document}